\documentclass[12pt,a4paper]{article}

\usepackage[affil-it]{authblk}
\usepackage{amssymb}
\usepackage{latexsym}
\usepackage{url}
\usepackage{xcolor}
\usepackage{hyperref}
\usepackage{multirow}
\usepackage{todonotes}
\usepackage[english]{babel}
\usepackage{blindtext}
\usepackage[a4paper]{geometry}
\usepackage{natbib}

\newgeometry{margin=0.85in}

\title{An automated near-real time computational method for induction and treatment of scar-related ventricular tachycardias}

\author[1]{Fernando O. Campos\thanks{Corresponding author: School of Biomedical Engineering and Imaging Sciences, Rayne Institute, 4th Floor, Lambeth Wing, St. Thomas' Hospital, Westminster Bridge Road, London SE1 7EH, United Kingdom. \textit{e-mail}: fernando.campos@kcl.ac.uk}}
\author[4]{Aurel Neic}
\author[1]{Caroline Mendonca Costa}
\author[1]{John Whitaker}
\author[1]{Mark O'Neill}
\author[1]{Reza Razavi}
\author[1]{Christopher A. Rinaldi}
\author[5]{Daniel Scherr}
\author[1]{Steven A. Niederer}
\author[2,3]{Gernot Plank\footnote{Joint senior authors.}}
\newcommand\CoAuthorMark{\footnotemark[\arabic{footnote}]}
\author[1]{Martin J. Bishop\protect\CoAuthorMark}

\affil[1]{\small{School of Biomedical Engineering and Imaging Sciences, King's College London, United Kingdom}}
\affil[2]{\small{Division of Biophysics, Medical University of Graz, Austria}}
\affil[3]{\small{BioTechMed-Graz, Austria}}
\affil[4]{\small{NumeriCor GmbH, Graz, Austria}}
\affil[5]{\small{Clinical Department of Cardiology, Medical University of Graz, Austria}}

\date{}

\UseRawInputEncoding

\begin{document}
\maketitle

\vspace{-1cm}

\begin{abstract}

Catheter ablation is currently the only curative treatment for scar-related ventricular tachycardias (VTs).
However, not only are ablation procedures long, with relatively high risk, but success rates are punitively low, with frequent VT recurrence.
Personalized \textit{in-silico} approaches have the opportunity to address these limitations.
However, state-of-the-art reaction diffusion (R-D) simulations of VT induction and subsequent circuits used for \textit{in-silico} ablation target identification require long execution times, along with vast computational resources, which are incompatible with the clinical workflow. 
Here, we present the Virtual Induction and Treatment of Arrhythmias (VITA), a novel, rapid and fully automated computational approach that uses reaction-Eikonal methodology to induce VT and identify subsequent ablation targets. 
The rationale for VITA is based on finding isosurfaces associated with an activation wavefront that splits in the ventricles due to the presence of an isolated isthmus of conduction within the scar;
once identified, each isthmus may be assessed for their vulnerability to sustain a reentrant circuit, and the corresponding exit site automatically identified for potential ablation targeting.
VITA was tested on a virtual cohort of 7 post-infarcted porcine hearts and the results compared to R-D simulations. 
Using only a standard desktop machine, VITA could detect all scar-related VTs, simulating activation time maps and ECGs (for clinical comparison) as well as computing ablation targets in 48 minutes. 
The comparable VTs probed by the R-D simulations took 68.5 hours on 256 cores of high-performance computing infrastructure.
The set of lesions computed by VITA was shown to render the ventricular model VT-free.
VITA could be used in near real-time as a complementary modality aiding in clinical decision-making in the treatment of post-infarction VTs. 

\end{abstract}

\textit{Keywords:} Myocardial Infarction, Ventricular Tachycardia, Catheter Ablation, Computational Modeling\makeatletter{\renewcommand*{\@makefnmark}{}
\footnotetext{Preprint submitted to Medical Image Analysis (Date: \today)}\makeatother}

\restoregeometry

\section{Introduction}

Ventricular tachycardia (VT), associated with myocardial infarction, is an important cause of sudden cardiac death~\cite{Koplan2009,El-Sherif2017}.
The only curative treatment for patients with incessant VT is radio-frequency ablation, involving the intense delivery of energy to key targets within the ventricle in order to disrupt the reentrant electrical excitation patterns that sustain the VT~\cite{Stevenson2013}.
This highly-invasive therapy is, however, associated with long procedure times and high complication rates, whilst success rates are punitively low; typically $>$\,50\% of patients will present with VT recurrence within 1-year post-procedure~\cite{Aliot2009,Tanawuttiwat2016,Willems2020}.

Failure of ablation therapy is most often due to an inability to comprehensively identify the myocardial substrate capable of sustaining the VT~\cite{Wissner2012}.
Attempts to reconstruct the complex intramural propagation pathways through the regions of structural remodelling to identify targets can be limited due to the surface-nature of catheter-based electroanatomical mapping (EAM) systems combined with inadequate spatial resolution~\cite{Graham2017}.
Furthermore, target identification is most accurate when VT is induced (and mapped) during the procedure which significantly increases the risk-profile and is often not haemodynamically-tolerated, and therefore not feasible~\cite{Stevenson2009}.
Importantly, multiple non-clinical VTs are frequently induced in a patient during an ablation procedure, with an inability to induce the arrhythmia responsible for the clinical presentation.
Ablation and termination of these induced VTs may have minimal impact on their clinical VT, and is thought to be another key reason behind the high recurrence rate post-ablation.

Recent advances in computational patient-specific modelling have demonstrated the initial feasibility of using \textit{in-silico} image-based models to perform virtual VT induction protocols~\cite{Arevalo2013,Arevalo2016}, identify reentrant circuits and subsequently target key ablation sites~\cite{Chen2016,Prakosa2018,Cedilnik2018}.
The entirely non-invasive nature of these approaches, requiring only magnetic resonance imaging (MRI) (or computed tomography - CT) imaging data to construct anatomical models, facilitates pre-procedure planning.
Furthermore, the ability to apply ``aggressive'' virtual VT-induction protocols, from numerous sites within the ventricles, allows multiple VTs to be probed, increasing the likelihood of inducing the clinical VT.
Once induced, detailed analysis of the ensuing circuits may be performed in conjunction with ``simulated ablation'' and subsequent re-induction testing, providing the opportunity to directly test target efficacy.
Studies using these approaches have demonstrated close association between the geometrical location of \textit{in-silico} ablation target predictions and clinical lesions associated with acute success in both pre-clinical and clinical retrospective studies~\cite{Ashikaga2013,Prakosa2018,Boyle2019}.
In addition, a proof-of-concept prospective study including 5 patients has been performed in which model-based predictions were used to directly guide the ablation procedure~\cite{Prakosa2018}.
Ablation lesions were created at locations suggested by simulations, with overall successful results.
Other studies using similar methods, not focusing directly on ablation, have demonstrated close agreement between personalized model predictions of VT circuits and exit sites with clinically measured VT activation maps during the procedure using both MRI- and CT-based models~\cite{Ashikaga2013,Chen2016,Deng2019B}.

Undoubtedly, these early successes have paved the way towards \textit{in-silico} personalized model guidance to play a leading role in VT ablation procedures in the (very) near future.
However, integral to full clinical translation necessitates the ability to fully integrate into a clinical workflow, both in terms of time-frames and the requirement for additional (computational) resources.
Current approaches largely require vast computational resources to perform monodomain simulations of cardiac electrophysiology (EP) that are costly due to the relatively high-resolution ($<$\,350\,$\mu$m) meshes required
to ensure adequate convergence of numerical solutions,
and the long-duration virtual induction protocols that need to be performed to identify all potential circuits.
Such requirements necessitate the use of off-site high-performance computing (HPC) infrastructure for simulating (still requiring many hours of compute time) and analyzing the many hundreds of gigabytes of data generated, prior to its use by a clinician~\cite{Arevalo2016,Prakosa2018}.

Eikonal models seek to simulate the spatio-temporal evolution of the wavefront of excitation with the important advantage of maintaining accuracy even when using much coarser computational meshes than required for full monodomain simulations~\cite{Colli-Franzone1990,Keener1991}.
The advent of Reaction Eikonal (R-E) approaches offer the potential of combining the advantages of an Eikonal model with the biophysical detail of reaction-diffusion (R-D) approaches, allowing accurate representation of upstroke morphology and repolarization dynamics~\cite{Jacquemet2012,Neic2017,Chen2016,Cedilnik2018}.
The R-E model thus provides an efficient representation of membrane dynamics that can be coupled with full- or pseudo-bidomain models to simulate extracellular potential distributions at a significantly reduced computational cost~\cite{Neic2017}.
The main current limitation of R-E approaches, however, is their ability to simulate only single, paced beats, \textit{i.e.} a single activation of all nodes within the domain.

In this study, we present an innovative \textit{in-silico} approach
that rapidly, and automatically, locates both vulnerable reentry circuits and corresponding ablation targets, in close to real-time using only standard desktop computational resources.
We term this non-invasive approach for planning VT ablation VITA (Virtual Induction and Treatment of Arrhythmias).
VITA creatively uses R-E infrastructure to identify isolated channels through infarcted regions, assesses their individual vulnerability to form a reentrant circuit, automatically identifies corresponding isthmus exits/entrances, and performs sequential virtual ablation of all target sites to output a minimal (optimal) set of ablation lesions.
Following a detailed description of the specific methodological detail underlying our approach, we directly test it against the current gold-standard virtual ablation targeting methodology, quantitatively comparing the VT circuits probed and corresponding computational requirements.

\section{Methods}

\subsection{Computational geometries}

Initially, we illustrate the individual steps in VITA using a bi-ventricular geometry of the human heart equipped with an idealized infarct scar.
Subsequently, VITA is applied to a virtual cohort of 7 infarcted porcine hearts.
Details of the geometrical models are given bellow.

\subsubsection{Idealized model}
\label{sect:H2C_model}

\begin{figure}[!t]
    \centering
    \includegraphics[width=0.85\textwidth]{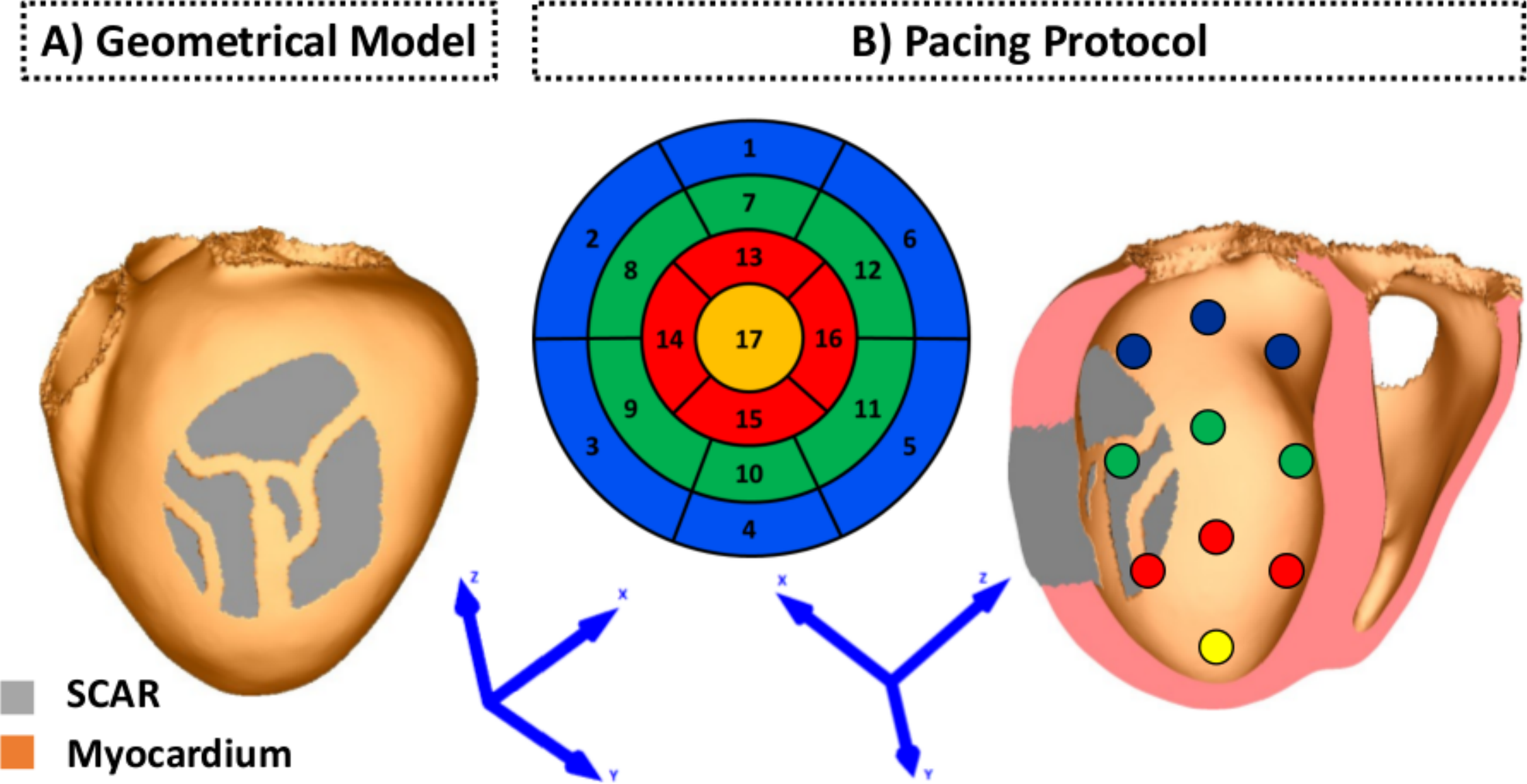}
    \caption{Idealized human bi-ventricular infarct model.
             A: Geometrical model of the human heart with an idealized infarct.
             Myocardial tissue is shown in orange and infarct scar in gray.
             B: Myocardial wall segments (numbered), as per the American Heart Association nomenclature, and associated pacing sites on the endocardial surface of the left ventricle (right panel).
            }
    \label{fig:H2C}
\end{figure}

The bi-ventricular model shown in Fig.~\ref{fig:H2C} was constructed from a whole heart, end diastolic, 3D, steady state free precession cardiovascular magnetic resonance (CMR) with an isotropic resolution of 1.3\,mm.
Tetrahedral finite elements (FEs) with a mean edge of 976\,$\mu$m discretization were used to construct the geometrical model~\cite{Prassl2009}.
In total, 2\,459\,832 elements and 464\,407 nodes make up the computational mesh.
Myocardial fiber orientations were incorporated into the model using a rule-based approach~\cite{Bayer2012}.
Further details about the construction of the anatomical model can be found elsewhere~\cite{Crozier2016}.
An idealized infarct scar transcended by a network of conducting isthmuses was assigned to the model in order to create pathways capable of sustaining electrical reentries.

\subsubsection{Porcine cohort}

\begin{figure}[!t]
    \centering
    \includegraphics[width=0.96\textwidth]{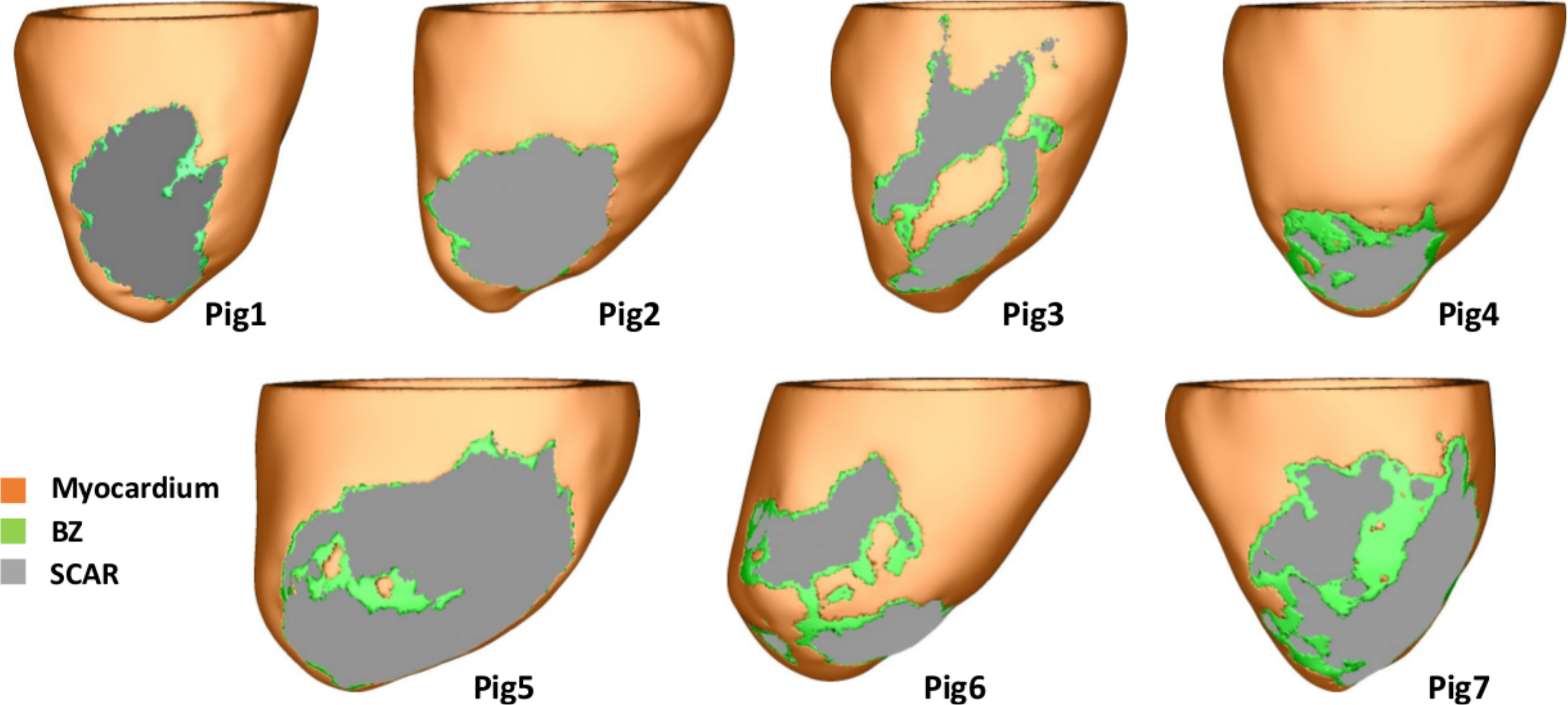}
    \caption{Cohort of seven porcine left ventricular anatomy models: healthy myocardium (orange), infarct BZ (green), and scar (gray).}
    \label{fig:Piggies}
\end{figure}

The virtual cohort of porcine left ventricular (LV) models was built based on late-gadolinium enhanced CMR imaging with an isotropic voxel resolution of 1\,mm obtained seven weeks following myocardial infarction~\cite{Whitaker2019}.
Details of the pipeline used in our group to convert CMR scans into geometrical models of the heart
have been reported previously~\cite{Campos2021}.
Briefly, the LV myocardium and blood pool were manually segmented by a single observer using Seg3D (www.seg3d.org).
The signal intensity range within the segmented myocardium was used to calculate the threshold for scar at 60\% of the maximum intensity within the image, and the threshold for the infarct border zone (BZ) tissue at 40\%~\cite{Perez-David2011}.
Tetrahedral FE meshes were created based on the segmented MRI data using custom written software based on the freely available library CGal~\cite{CGal2018}.
The resulting high resolution models shown in Fig.~\ref{fig:Piggies} were discretized with an average FE edge length of 338\,$\mu$m~\cite{Prassl2009}.
The meshes in Fig.~\ref{fig:Piggies} were down-sampled to 888\,$\mu$m with Meshtool~\cite{Neic2020} and used to test VITA.
As in the idealized model, a rule-based approach was employed to assign fiber orientations within the myocardium~\cite{Bayer2012}.

\subsection{Eikonal model}

Eikonal models sit at the core of VITA as they represent a fast way to simulate the cardiac
wavefront of excitation using much coarser computational meshes~\cite{Colli-Franzone1990,Keener1991}.
Activation wavefronts are computed by solving the Eikonal equation:
\begin{eqnarray}
    \sqrt{\nabla t_{\rm a}^\top \, \mathbf{V} \, \nabla t_{\rm a}} & = & 1 \qquad \textit{in } \Omega \\
    t_{\rm a} & = & t_0 \qquad \textit{in } \Gamma
    \label{eq:Eikonal}
\end{eqnarray}
where $t_{\rm a}$ is a function describing the wavefront arrival time at location $\mathbf{x}$,
$\mathbf{V}$ is the squared conduction velocity (CV) tensor, which combines with the longitudinal and
transversal fiber orientations $(l, t)$ in the myocardium $\Omega$
with associated scalar velocities $(v_{l}, v_{t})$.
The boundary condition $t_0$ in $\Gamma$ represents location and timing of the stimulation
initiating activation.
CVs of $v_{l}$ = 0.50\,m/s and $(v_{t}$ = 0.16\,m/s were assigned to the healthy myocardium
to reproduce CVs obtained in biophysical simulations~\cite{MendoncaCosta2013}
using conductivities values reported in the literature~\cite{Hooks2007}.

For computing distance fields, the Eikonal equation can be simplified by using
$v_l = v_t = 1$ and therefore $\mathbf{V} = \mathbf{I}$, the identity matrix, to
\begin{eqnarray}
    \| \nabla t_{\rm a} \| & = & 1 \qquad \textit{in } \Omega \\
    t_{\rm a} & = & 0 \qquad \textit{in } \Gamma
\end{eqnarray}
where $t_{\rm a}$ now represents the distance and travel time from $\Gamma$ to $\mathbf{x}$.

\subsection{Computational environment}
\label{sect:comput_environment}

All experiments using VITA were performed in coarser geometrical models and carried out on a 12-core Intel(R) Xeon(R) CPU E5-2670 v3 2.30\,GHz (64\,bit Linux) desktop machine with 64\,GB of RAM memory.
Biophysically realistic simulations within high resolution models were performed on 4 nodes of our internal HPC cluster machine, with each node consisting of a 64-core (2 sockets, 32 cores/socket) AMD EPYC (64\,bit Linux) machine equipped with 256\,GB of RAM memory.
Execution times are presented as the time each approach required to run on its respective computational environment as well as on a single CPU.
CPU times are taken as the execution time on the parallel computing environment multiplied by the total number of cores used to run each approach.

\subsection{Rationale of the method}

\begin{figure}[!t]
    \centering
    \includegraphics[width=0.60\textwidth]{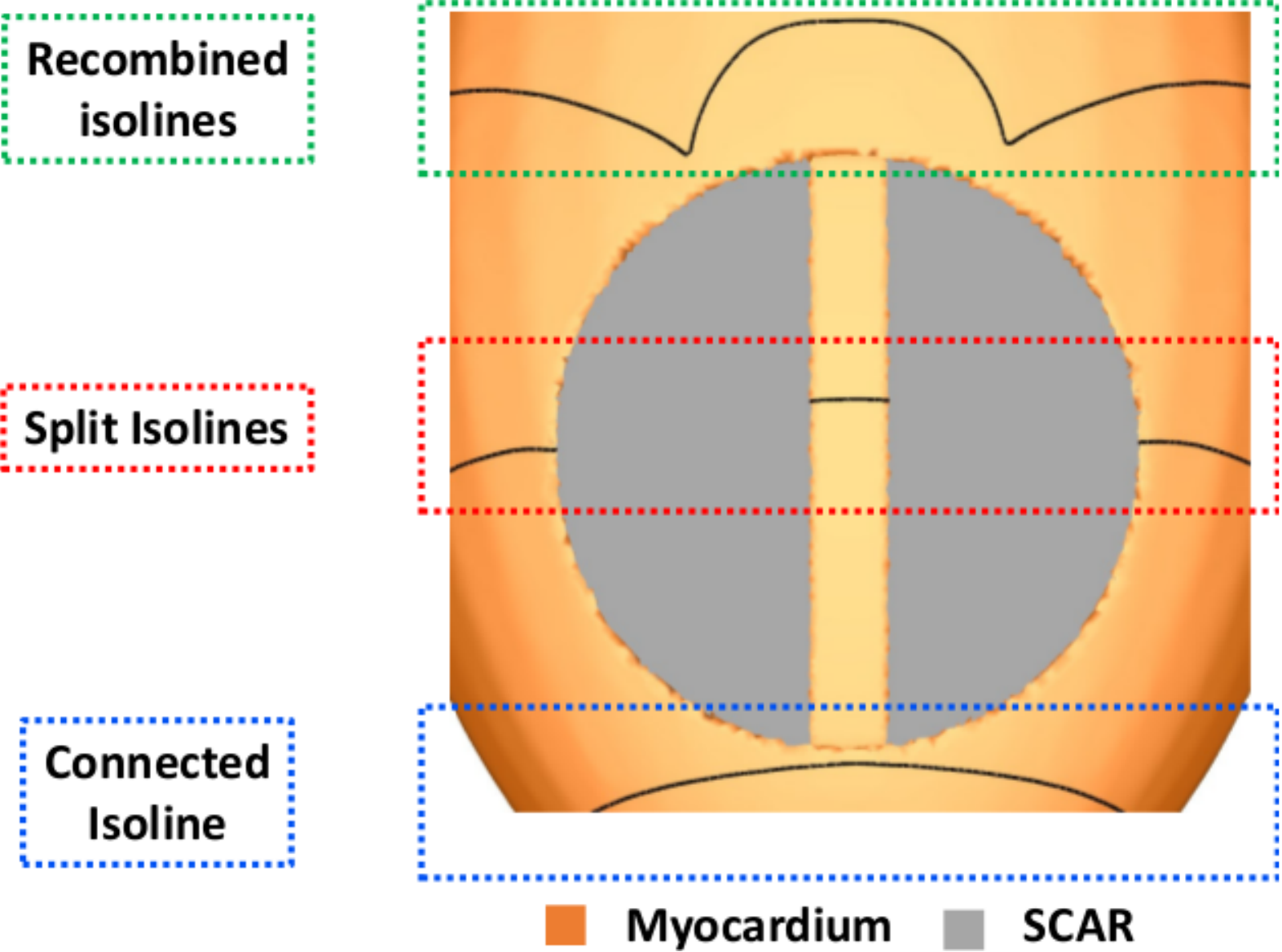}
    \caption{Splitting of isosurfaces.
             Isolines illustrating the regions where the wavefront splits due to the presence of an isthmus inside the scar (red box).
            }
    \label{fig:Split_surf}
\end{figure}

The rationale for the algorithm behind VITA is based on finding isosurfaces associated with a paced activation wavefront that splits within the ventricles due to the presence of isolated isthmuses generated by scar and surviving tissue.
Such isthmuses provide isolated conduction pathways that are disconnected from the rest of the myocardium.
This rationale is illustrated in Fig.~\ref{fig:Split_surf}.

In the absence of an isolated isthmus, a single connected isosurface exists (blue box); when it encounters the isthmus, the isosurface ``splits'' into two components (red box), with a smaller isosurface within the isthmus itself, which is unconnected to the larger isosurface wrapping around the rest of the ventricles due to the nonconducting scar at its boundaries.
By locating the regions in space at which these isolated isthmuses appear and disappear allows the identification of entry and exit sites, respectively, which could be used to guide ablation to target the required sites within the isthmus.

\begin{figure}[!t]
    \centering
    \includegraphics[width=0.65\textwidth]{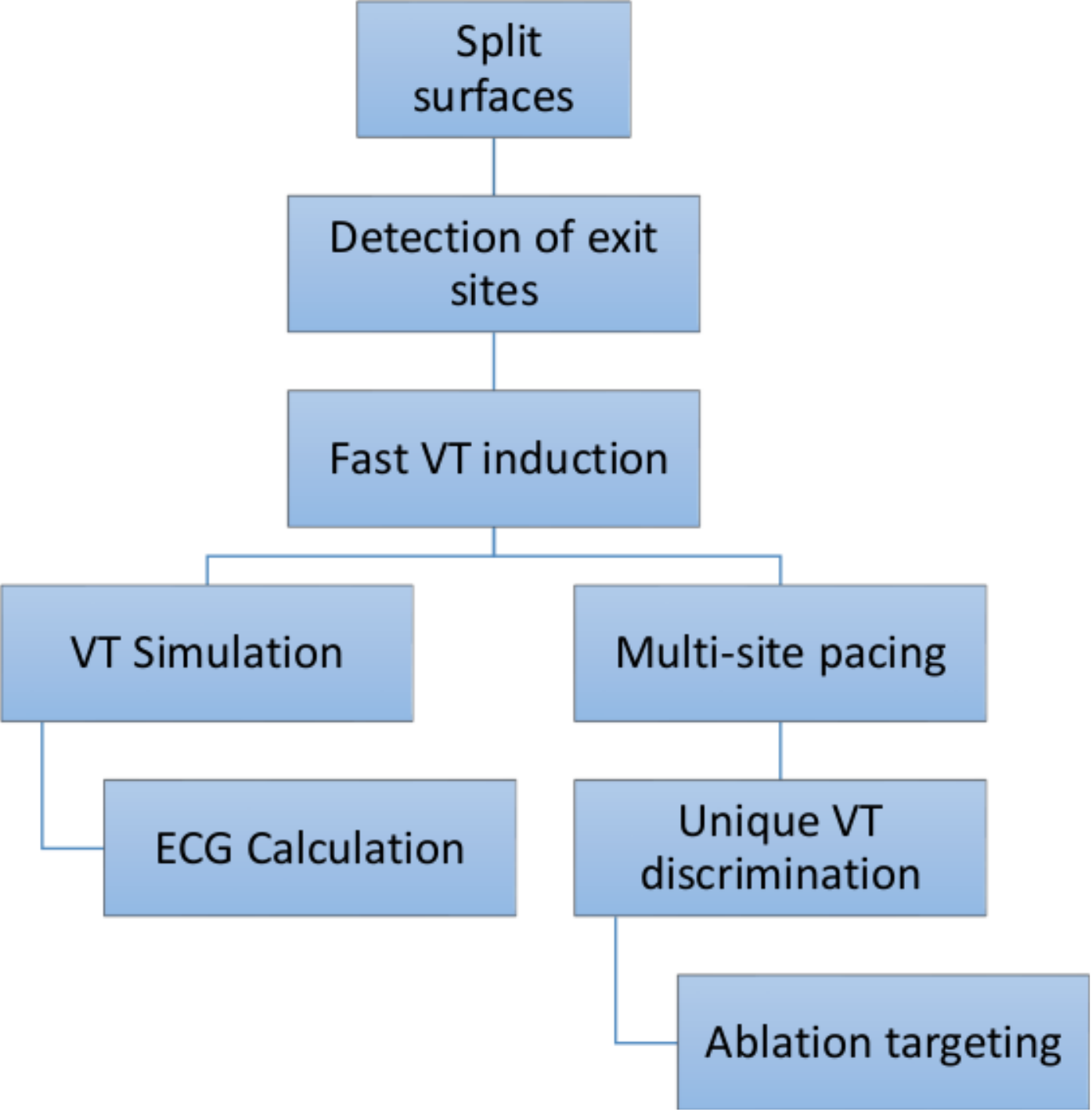}
    \caption{Conceptual diagram showing the steps of VITA.}
    \label{fig:VITA}
\end{figure}

Once an exit (or entry) site is located, the mesh is split along its respective isosurface to create a local region of unidirectional conduction block.
Cardiac activation is then simulated using a fast Eikonal model.
Fig.~\ref{fig:VITA} summarizes all steps comprising the VITA approach.
The specifics of each step is described in detail as follows.

\subsubsection{Split surfaces}
\label{sect:split_surfs}

\begin{figure}[!t]
    \centering
    \includegraphics[width=0.85\textwidth]{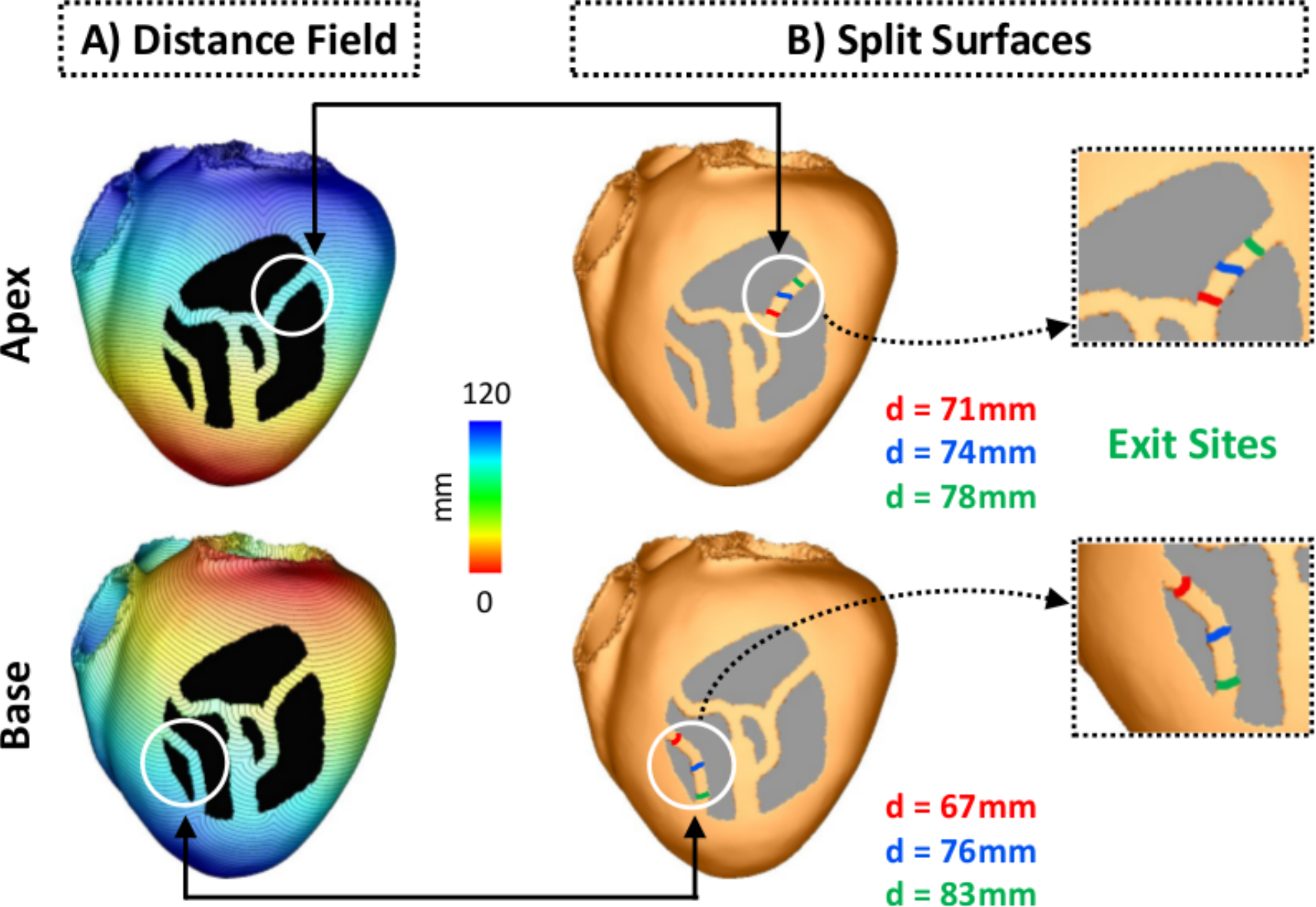}
    \caption{Detection of exit sites.
             A: Examples of distance fields computed from pacing sites located at the apex (upper panel) and base (bottom panel).
             B: Split surfaces detected within the isthmuses at different distances $d$ from the pacing sites.
             Isosurfaces in green are associated with the distal mouths, \textit{i.e.}, exit sites, of the isthmuses.
             3D surfaces are shown as lines on the LV epicardium for visualization purposes.
            }
    \label{fig:H2C_split_surfs}
\end{figure}

First, a distance field is computed from a source point, or stimulus site.
The distance field depends on the location of the stimulus and on the anatomy of the 3D scar.
Fig.~\ref{fig:H2C_split_surfs}A shows distance fields computed from pacing sites located at the LV apex (top panel) and base (bottom panel).
In both examples, isosurfaces are separated by 1\,mm (approximately the mean discretization of the geometrical model, see Sect.~\ref{sect:H2C_model}).
Locations where isosurfaces split within the LV are illustrated in Fig.~\ref{fig:H2C_split_surfs}B.
While both isosurfaces are seen as lines on the epicardial surface of the LV, they are in fact three-dimensional across the ventricular wall.
In total, 70 and 66 split isosurfaces were detected in the apical and basal distance fields, respectively.

\subsubsection{Detection of exit sites}
\label{sect:exit_sites}

Second, amongst all split isosurfaces only those without downstream neighbors are selected.
These represent regions where the specific isosurface terminates, by either recombining with a larger isosurface or by annihilation with a wavefront traveling in opposite direction (collision).
Such isosurfaces represent distal mouths or the exit sites of channel isthmuses within the scar.
Fig.~\ref{fig:H2C_split_surfs}B illustrates the detection of two different exit sites in the idealized model after pacing from the apex (top panel) and base (bottom panel).
Split surfaces detected within the isthmuses are shown in different colors based on their distances $d$ to the pacing sites.
Split surfaces in green are associated with distal mouths, \textit{i.e.}, exit sites (Fig.~\ref{fig:H2C_split_surfs}B, inset) as they do not have a downstream neighbor.
Split surfaces in red, on the other hand, represent entrances which are regions where the first split was detected (no upstream neighbor).
Note that the detected exit sites are not exactly located at the distal mouths of the isthmuses.
This is due to collisions of isosurfaces at those regions.
Although the detection of only one exit per stimulus site is illustrated in Fig.~\ref{fig:H2C_split_surfs}, in total, 3 (out of 70 split surfaces, see Sect.~\ref{sect:split_surfs}) and 2 (out of 66 split surfaces) exits were detected when pacing from the apex and base, respectively.
Under a given pacing location the exit sites, rather than entrances, were chosen as they are often used to guide catheter ablation therapy of VTs.

\subsubsection{Fast VT induction}
\label{sect:Eikonal}

\begin{figure}[!t]
    \centering
    \includegraphics[width=0.85\textwidth]{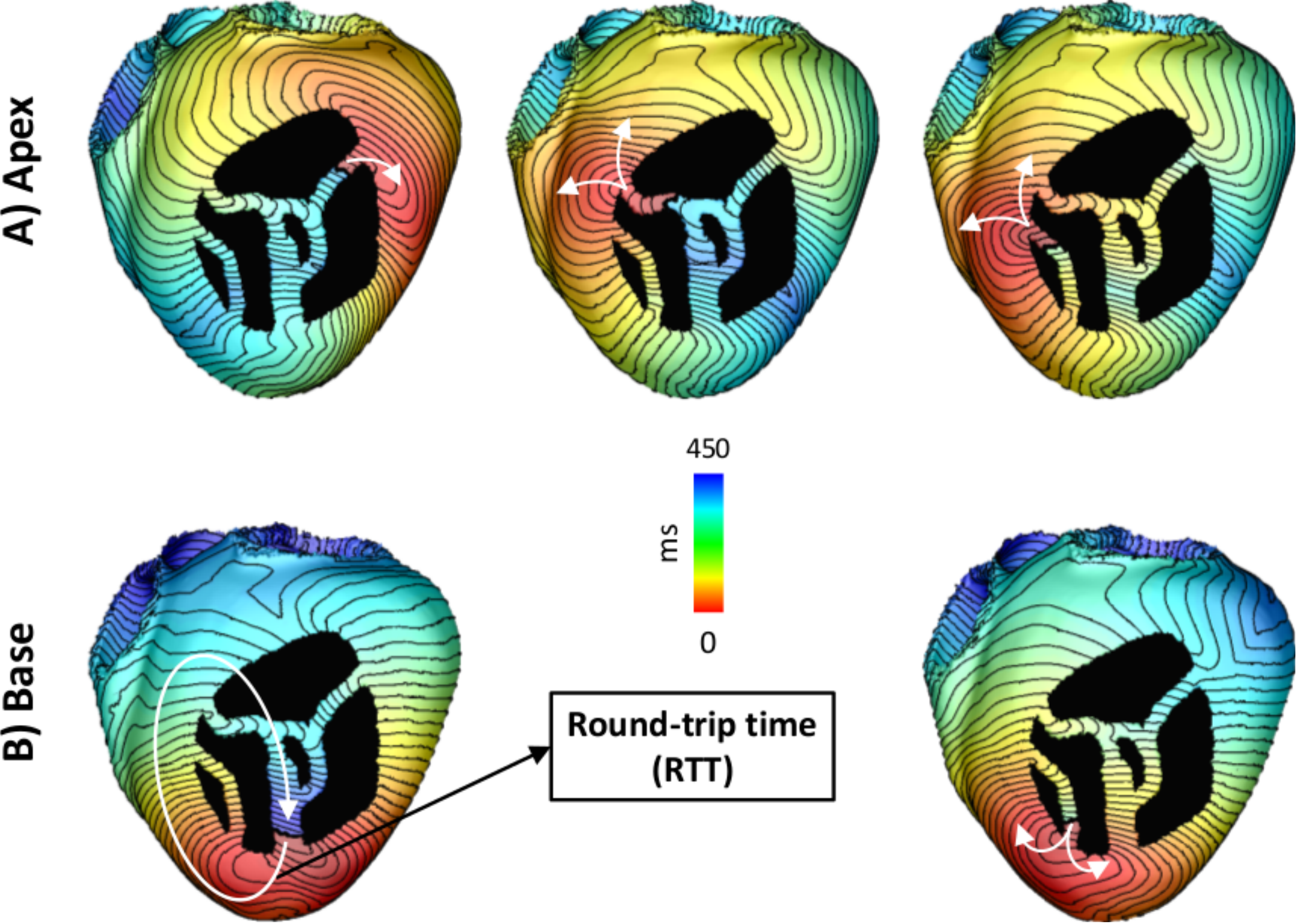}
    \caption{Pacing locations and induced VTs in the idealized  model.
             VTs associated with exit sites detected after pacing the model at the A) apex (upper panel) and B) base (lower panel).
             The round-trip time is the time the wavefront takes to travel back to the region of block.
            }
    \label{fig:H2C_uVTs_STIM}
\end{figure}

Once all exit sites for a given distance field are detected, the next step is to check whether they can sustain VTs.
This is achieved by simulating unidirectional conduction block at the exit site.
Unidirectional block was created by disconnecting the isosurface associated with the detected exit site, creating a region of uncoupled tissue across which activation cannot travel.
Isosurface disconnection is done using a discontinuous FE approach that imposes electrical insulation along edges of FEs~\cite{MendoncaCosta2014}.
Briefly, elements lying on the isosurface are marked for insulation.
Insulation is enforced by decoupling nodes shared between adjacent FEs.
Shared nodes are then duplicated, with one set located on one side of the insulating surface while the other set, with same spatial coordinates, is assigned at the opposite side of the isosurface.
Cardiac activation is then initiated on a set of nodes on one side of the decoupled surface, automatically ensuring unidirectional block of the wavefront, which propagates along a single direction along the isthmus.
Fig.~\ref{fig:H2C_uVTs_STIM} shows activation time (AT) maps of VTs associated with 3 and 2 exit sites detected after pacing the model at the apex (top panel) and base (bottom panel), respectively.
Activation was initiated by taking $t_0$ = 0\,ms (Eq.~\ref{eq:Eikonal}) at nodes located at the exit site.

The time the wavefront takes to travel back to the region of block, \textit{i.e.} the activation time of the nodes on the distal side of the decoupled surface (see Fig.~\ref{fig:H2C_uVTs_STIM}), is referred here as the round-trip time (RTT).
RTT is used here as a metric to assess whether the reentrant pathway associated with a detected exit site is long enough to sustain VTs.

\subsubsection{Simulating VT dynamics and ECGs}
\label{sect:R-E}

VITA also allows for a quick investigation of electrical wavefront propagation during VT patterns,
and associated extracellularly sensed potential fields,
either sensed as electrocardiogram (ECG) at the body surface or as electrogram (EGM) by an implanted device.
This extra step is achieved by solving a R-E model that uses eikonal-based activation patterns to compute the distribution of electric sources~\cite{Neic2017}.
Unlike R-D models, R-E models are not locked into into using high spatial resolutions,
and, as such, can be solved quickly, with close to real-time performance.
The reaction part of both R-E and R-D models is simulated by solving a computational model of the ventricular action potential (AP) such as the ten Tusscher (TT) model~\cite{tenTusscher2006}.

\begin{figure*}[!t]
    \centering
    \includegraphics[width=\textwidth]{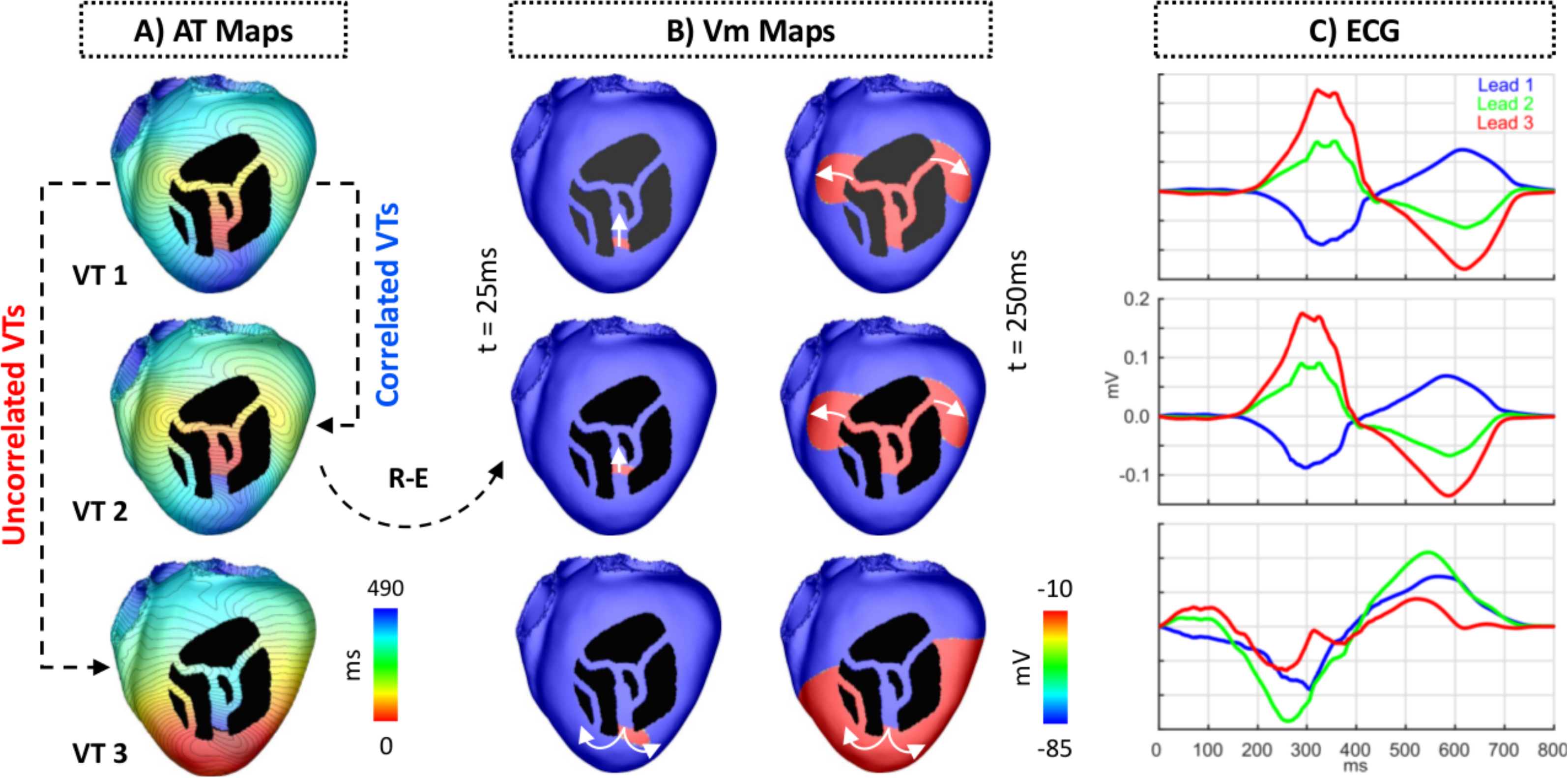}
    \caption{VT and ECG simulation using the R-E model.
             A: AT maps of three VTs induced in the idealized model.
             VT\,1 and VT\,2 are correlated as they share the same circuit and have the same chirality.
             VT\,1 and VT\,3 are not correlated because they have opposite chirality despite sharing the same circuit.
             B: Snapshots of the transmembrane potential $V_m$ generated by the R-E model at times t = 25\,ms (left column) and t = 250\,ms (right column).
             White arrows represent the direction of propagation.
             C: ECGs associated with each VT in A.
             Shown are traces from leads 1, 2 and 3 of the Einthoven's triangle.
            }
    \label{fig:H2C_VTs_ECGs}
\end{figure*}

The R-E model provides a fast way to obtain EGM and ECG signatures of the VT for clinical comparison.
ECGs are obtained by sampling the potential field at lead locations that approximate clinical standard electrode positions.
This can be achieved, with high fidelity, using a lead field approach~\cite{Potse2018,Gillette2021},
or, approximately, by resorting to $\phi_{\mathrm e}$ recovery techniques that make the simplifying assumption of the heart being immersed in a uniform volume conductor of infinite extent~\cite{Plonsey1998}.
Examples of computed ECGs related to the Einthoven limb leads and their associated VTs are shown for one VT cycle in Fig.~\ref{fig:H2C_VTs_ECGs}.

R-E and R-D simulations within all models presented here were performed with CARPentry-Pro~\cite{Neic2017} (NumeriCor GmbH, Graz, Austria), built upon openCARP~\cite{Plank2021}.
Algorithms underlying the R-D model, as implemented in the Cardiac Arrhythmia Research Package (CARP),
have been described extensively elsewhere~\cite{Vigmond2008} (http://carp.medunigraz.at).

\subsubsection{Multi-site pacing}

As shown in Fig.~\ref{fig:H2C_uVTs_STIM}, different VTs can be induced depending on the distance field, or pacing location in a clinical setting.
This suggests that multi-site pacing may be necessary to detect all vulnerable regions.
Multi-site pacing was implemented by computing distance fields from the 17 locations on the endocardial surface of the LV shown in Fig.~\ref{fig:H2C}B.

In total, 53 VTs were detected by VITA for all 17 pacing locations in the idealized model.
All VTs satisfied two conditions: they were associated with an exit site and had a RTT of at least 50\,ms.
While this value is lower than reported VT cycle lengths~\cite{Nishimura2021}, it allowed us to highlight as many circuits as possible, including less likely VTs that may only form under more extreme pathological conditions
such as very short AP duration (APD) and very slow conduction within the isthmus network.

\subsubsection{Unique VT discrimination}

\begin{figure*}[!t]
    \centering
    \includegraphics[width=\textwidth]{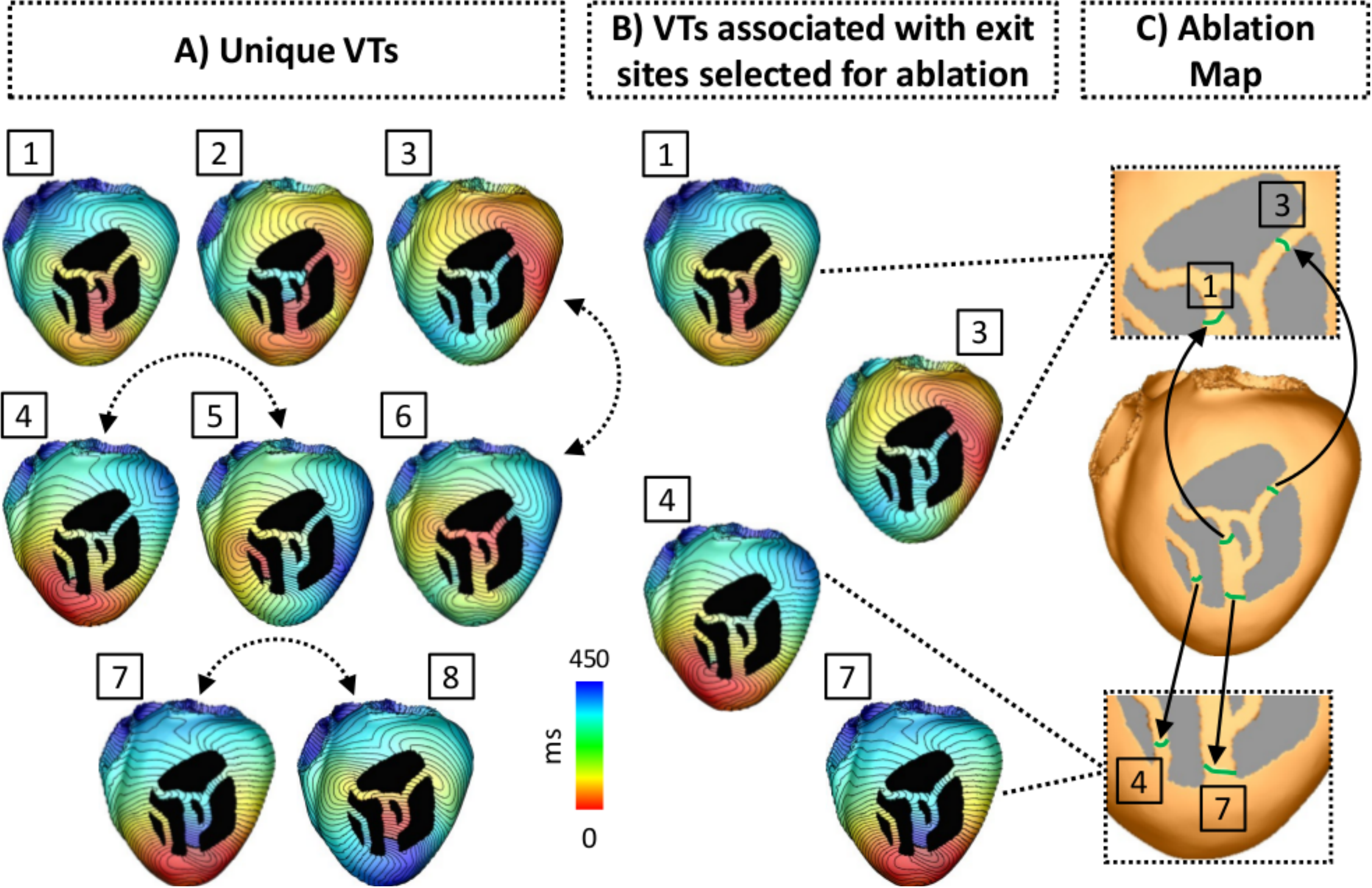}
    \caption{Unique VTs and ablation targeting in the idealized model.
             A: Unique VTs detected by VITA after pacing the bi-ventricular model from all 17 locations shown in Fig.~\ref{fig:H2C}B.
             In total, 8 unique VTs were detected.
             Dashed arrows represent VTs induced from similar exit sites, but have opposite directions.
             B: VTs associated with exit sites selected for ablation.
             Among all unique VTs, those associated with exit sites 1, 3, 4 and 7 were optimally selected for ablation.
             C: Optimal ablation map output by VITA.
             Ablation lesions are shown in green.
             Insets show the zoomed-in split surfaces associated with exit sites marked for ablation.
            }
    \label{fig:H2C_uVTs_ABL}
\end{figure*}

Several VTs induced by the VITA protocol may be similar or even identical.
This is illustrated in Fig.~\ref{fig:H2C_VTs_ECGs}A, which shows three VTs induced from separate exit sites.
In a next step identical or very similar VTs will be reduced to a set of unique VTs.
Two approaches were considered.
First, VT uniqueness was assessed by computing the correlation coefficient between QRS complexes of the simulated ECG signals linked to a given VT (Fig.~\ref{fig:H2C_VTs_ECGs}C).
This approach follows EAM procedures where VT circuits are identified by comparing the QRS signature of a VT to a template QRS signature of a previously recorded VT.
While this approach offers the potentially significant advantage of comparing QRS signatures to available EGM or ECG signals to ascertain that simulated and observed VT are closely related, the power to robustly discriminate VTs of subtly different reentrant pathways was limited.
Thus, secondly, correlation coefficients were also computed between AT maps to assess uniqueness among very similar VTs.
For this sake, AT maps were cyclically aligned prior to the correlation calculation based on the RTT, such that they had the same starting location ($t$ = 0).
The Pearson product-moment was used to compute the correlation coefficient between AT maps of two VTs.
Different threshold values for the correlation coefficients were tested for the detection of unique VTs.
VTs with a correlation coefficient smaller than (or equal to) the threshold were considered unique, whereas VTs with coefficients larger than that were considered redundant.
The threshold of 0.8 was chosen as values above it resulted in redundant VTs.
All unique VTs detected by VITA in the idealized model are shown in Fig.~\ref{fig:H2C_uVTs_ABL}A.

Note that there are VTs induced from similar exit sites (dashed arrows), but these VTs have opposite chirality and are considered here to be unique (correlation coefficient smaller than the threshold).
Also, the exit sites associated with VTs 1 and 2 are located inside the isthmus, rather than at one of the mouths connecting the isthmus to the bulk LV.
This is because of 1) distance fields associated with pacing sites located on the opposite side of the heart (in relation to the infarct) led to collisions of isosurfaces within the isthmus.
Such isosurfaces were classified as exit sites as they met the criterion of not having downstream neighbors;
and 2) the anatomy of the idealized infarct.
As can be seen Fig.~\ref{fig:H2C_uVTs_ABL}, surviving bundles of myocardium branch and merge within the scar (larger central isthmus). This led to the formation of exit sites near those regions.

\subsubsection{Ablation targeting}
\label{sect:ABL-targeting}

The last step of VITA is the detection of ablation targets aiming to terminate all VTs.
The detection of ablation targets is performed by iterating over the ordered list of exit sites.
Starting with the exit site of longest RTT, a unidirectional block is imposed by disconnecting the nodes associated with the exit site as described in Sects.~\ref{sect:split_surfs}-\ref{sect:exit_sites} and the associated VT is triggered by initiating propagation at the proximal face of the block.
If wavefront arrival is detected at the opposite face of the temporal block, the exit site is virtually ablated, \textit{i.e.} the block is made permanent.
The procedure is repeated for each exit site in the ordered list.
The resulting list of selected exit sites form the optimal ablation map shown in Fig.~\ref{fig:H2C_uVTs_ABL}C.

\subsection{Comparison to current approaches}
\label{sect:R-D}

Following \cite{Arevalo2013,Ashikaga2013,Arevalo2016,Pashakhanloo2018,Prakosa2018,Deng2019A,Deng2019B},
biophysically detailed simulations of VT induction were performed to serve as a basis for comparison
with the results obtained with VITA.
The monodomain equation was used to simulate the electrical activity within high resolution porcine cohort:

\begin{eqnarray}
    \nabla \cdot (\mathbf{\sigma_\mathrm{m}} \nabla V_\mathrm{m}) &=&  \beta I_\mathrm{m},
    \label{eq:monodomain1}    \\
    C_\mathrm{m}\frac{\partial V_\mathrm{m} }{\partial t} + I_{\mathrm{ion}} (V_\mathrm{m},\mathbf{\eta}) - I_{\mathrm{stim}} &=& I_\mathrm{m},
    \label{eq:monodomain2}    \\
    \frac{\partial \mathbf{\eta}}{\partial t} &=& f(V_\mathrm{m},\mathbf{\eta})
    \label{eq:monodomain3}
\end{eqnarray}
where $V_\mathrm{m}$ is the transmembrane voltage;
$\beta$ is the bidomain surface to volume ratio;
$I_\mathrm{m}$ is the transmembrane current density;
$C_\mathrm{m}$ is the membrane capacitance per unit area;
$I_{\mathrm{ion}}$ is the density of the total ionic current flowing through the cell membrane;
$I_{\mathrm{stim}}$ is the stimulus current density;
and $\mathbf{\sigma_\mathrm{m}}$ is the harmonic mean conductivity tensor with eigenvalues given as:

\begin{equation}
    \mathbf{\sigma_\mathrm{m}} = \sigma_i^\zeta(\sigma_i^\zeta + \sigma_e^\zeta)^{-1}\sigma_e^\zeta
    \label{eq:harmonic_mean}
\end{equation}
where $\zeta = l\:|\:t$ are the eigendirections of the tissue along the cardiac fiber direction ($\zeta = l$) and transverse ($\zeta = t$) to it~\cite{Bishop2010}.

Equations were solved using CARPentry which is able to match exactly monodomain simulation results as obtained with the legacy CARP simulator~\cite{Vigmond2003,Vigmond2008} used in the reference studies~\cite{Arevalo2013,Ashikaga2013,Arevalo2016,Pashakhanloo2018,Prakosa2018,Deng2019A,Deng2019B}.
Temporal discretization used an implicit-explicit Crank-Nicolson scheme with operator splitting at a time step of 25\,$\mu$s~\cite{Niederer2011}.
Mass lumping was turned off.
Detailed numerical settings are reported in the supplement to enable replication of simulation results with openCARP~\cite{Plank2021}.

\subsubsection{Electrophysiological parameters}

Anisotropic conductivity values of $\sigma_m^l$ = 0.1361\,S/m and $\sigma_m^t$ = 0.0176\,S/m were assigned to the myocardium based on intra- and extracellular values reported in the literature~\cite{Hooks2007}.
The TT model~\cite{tenTusscher2006} was used to simulate channel gating and ionic concentrations in cardiac cells
(represented by $\eta$ in Eq.~\ref{eq:monodomain3}).
The maximum conductances of the rapid ($I_{Kr}$) and slow ($I_{Ks}$) delayed rectifier currents were increased
by factor of 2.3 to  shorten APD.
The resulting APD of 201.5\,ms (basic cycle length of 500\,ms) falls within the values reported in optical recordings in the porcine heart~\cite{Kong2007}.

Furthermore, modifications to the TT model were implemented to represent EP remodelling in the infarct BZ~\cite{Arevalo2013,Ashikaga2013,Arevalo2016,Pashakhanloo2018,Prakosa2018,Deng2019A,Deng2019B}.
Specifically, the maximum conductances of fast sodium current ($I_{Na}$) and L-type calcium were reduced
to 38\% and 31\% of their control values, respectively.
Also, $I_{Kr}$ and slow $I_{Ks}$ were reduced by 70\% and 80\%, respectively.
These changes resulted in longer APD in the BZ when compared to the healthy myocardium.

Isotropic tissue conductivity with values of $\sigma_{ml}$ = $\sigma_{mt}$ = 0.064\,S/m~\cite{Luke1991} were assigned to the BZ
in order to account for structural remodelling such as fiber disarray and presence of fibrosis~\cite{MendoncaCosta2018}.
The scar was represented as being nonconducting (\textit{i.e.},
by imposing no-flux boundary condition at its interface, sealing the intracellular space there).

\subsubsection{Pacing protocol}

VTs were induced in the high resolution porcine cohort following the Virtual-heart Arrhythmia Risk Predictor (VARP)~\cite{Arevalo2016} and Virtual-heart Arrhythmia Ablation Targeting (VAAT)~\cite{Ashikaga2013,Prakosa2018} approaches:
a clinical S1-S2-S3 electrical stimulation protocol was applied from 17 pacing sites on each pig model as illustrated in Fig.~\ref{fig:H2C}B.
This virtual multi-site pacing allows the investigation of a wide range of possible VT morphologies that could arise from a given infarct morphology.
For simplicity we will now only refer to this approach as VARP.

Prior to organ-scale pacing, the LV models were initialized with single-cell model states obtained after pacing the TT model for 100 cycles with a basic cycle length of 600\,ms.
In each LV model, 8 stimuli (S1) at a cycle length of 600\,ms were delivered,
followed by a premature stimulus (S2) delivered 250\,ms after the last S1 beat.
The coupling interval of 250\,ms was chosen as it was the shortest interval allowing capture in tissue.
If the S2 beat did not result in reentry, an additional S3 stimulus was delivered 180\,ms (shortest interval required for capture) after the S2~\cite{Arevalo2016,Pashakhanloo2018,Prakosa2018}.
In all simulations, the virtual pacing electrode had a shape of a half sphere with center at each of the 17 segments in Fig.~\ref{fig:H2C}B and 1\,mm radius.
Stimulation was performed by injecting a transmembrane current $I_{\mathrm{stim}}$ as in Eq.~\ref{eq:monodomain2}.
Simulations were monitored for 2\,s after the last paced beat to detect the presence of VTs.
Similar to previous studies~\cite{Arevalo2016,Pashakhanloo2018},
VTs lasting until the end of the 2\,s-window were considered sustained.
Detection of unique VTs obtained with the VARP approach was performed by visual inspection of wavefronts.

\subsection{Simulation of ablation lesions}

\textit{In-silico} ablation was performed according to the ablation maps output by VITA (see Fig.~\ref{fig:H2C_uVTs_ABL}).
The maps were interpolated from the down-sampled to the high resolution models using Meshtool~\cite{Neic2020}.
The ablation maps represent a very small (optimal) set of targets, which do not aim to replicate clinical lesion dimensions as in previous studies~\cite{Prakosa2018}.
Ablation lesions were modelled by rendering the tissue (FE elements) at the target areas nonconducting as the necrotic scar.
Both VITA and VARP induction protocols as described above were then repeated to assess whether VT could still be induced.

\section{Results}

The VITA approach presented in the previous sections was tested in the virtual cohort of porcine LV models.
The number of arrhythmias induced with VITA was compared to those generated after applying the VARP protocol to the high resolution LV models.
Next, a sensitivity analysis was done on VITA.
Finally, ablation maps were constructed based on the detected unique VTs.

\subsection{VT inducibility}

Table~\ref{tab:uVTs} summarizes the number of VTs detected after applying both VITA and VARP approaches to each porcine LV model.
Using VITA, VTs were inducible in Pigs 3, 5, 6 and 7.
Although numerous VTs were induced in the same model, a subset of these VTs were similar as the same exit site was detected from different pacing locations (see Fig.~\ref{fig:H2C_VTs_ECGs}A for details).
In Pig 3, for instance, 46 VTs were induced, however, only 7 were unique.
Overall, more unique VTs were induced with VITA than with VARP.
For instance, 7 unique VTs were induced in Pig 3 with VITA while only one was induced using VARP (see Table~\ref{tab:uVTs}).
A fundamental difference in inducibility between VITA and VARP was witnessed
in pigs 4 and 5 where only either VITA or VARP induced a VT, but not both.

\begin{table}[!t]
    \centering
    \caption{Number of VTs induced within the virtual porcine cohort.}
    \begin{tabular}{|c|c|c|c|c|}
        \hline
        \multicolumn{ 1}{|c|}{\multirow{2}{*}{\textbf{Pig}}} & \multicolumn{ 2}{c|}{\textbf{Induced VTs}} & \multicolumn{ 2}{c|}{\textbf{Unique VTs}} \\ \cline{ 2- 5}
        \multicolumn{ 1}{|c|}{} & \textbf{VITA} & \textbf{VARP} & \textbf{VITA} & \textbf{VARP} \\ \hline
        1 & 0  & 0  & 0 & 0 \\ \hline
        2 & 0  & 0  & 0 & 0 \\ \hline
        3 & 46 & 15 & 7 & 1 \\ \hline
        4 & 0  & 1  & 0 & 1 \\ \hline
        5 & 5  & 0  & 1 & 0 \\ \hline
        6 & 97 & 12 & 5 & 4 \\ \hline
        7 & 75 & 7  & 6 & 3 \\ \hline
    \end{tabular}
    \label{tab:uVTs}
\end{table}

\subsection{Similarity in induced VTs}

\begin{figure*}[!t]
    \centering
    \includegraphics[width=\textwidth]{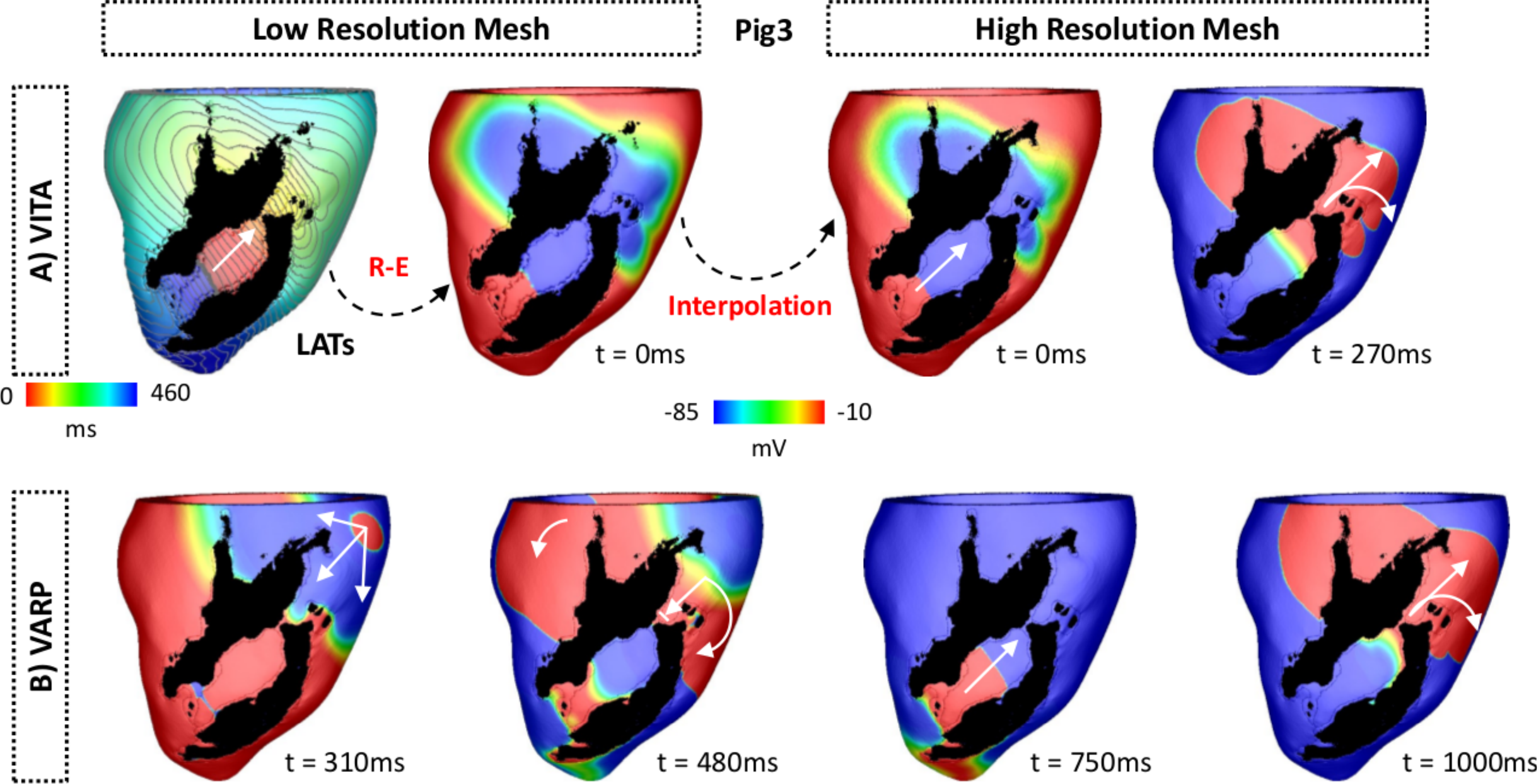}
    \caption{VT initiation in  Pig 3 ventricular model.
             A: Induced VT using VITA.
             Top left panel: AT map used to induce VT with the R-E solver within the down-sampled LV model.
             Top right panel: $V_\mathrm{m}$ map interpolated from the low to the high resolution model.
             $V_\mathrm{m}$ map obtained with the R-D model at t = 270\,ms is also shown.
             Propagation occurs unidirectionally from the exit site towards the base of the LV.
             The wavefront travels around the scar and reenters the isthmus from the apex.
             B: Induced VT using VARP.
             $V_\mathrm{m}$ maps at different times show VT induction following the S1-S2 pacing protocol.
             White arrows represent successful propagation and lined arrows represent conduction block.
             The S2 beat was blocked at the isthmus's mouth proximal to the stimulus site (t = 480\,ms) and, similar to the VT in A, the wavefront revolves around the scar reentering the isthmus (t = 750\,ms).
             VT induction in the LV model with both VITA and VARP can be fully appreciated in Supplemental Videos 1 and 2, respectively.
            }
    \label{fig:Piggy21_VT}
\end{figure*}

Fig.~\ref{fig:Piggy21_VT} shows VTs induced in Pig 3 using VITA (top panel) and VARP (bottom panel).
In VITA, the AT map of the detected VT was fed into the R-E solver and simulated for one cycle (see Sect.~\ref{sect:R-E} for details).
Both AT and $V_\mathrm{m}$ maps show the wavefront propagating in the apico-basal direction.
All state variables were then interpolated onto the high resolution mesh in Fig.~\ref{fig:Piggy21_VT}A, top right panel.
After leaving the isthmus the wavefront travels around the scar towards the apex (t = 270\,ms) where it can reenter, forming the circuit sustaining the VT.
The VT induced with VITA within Pig 3 can be fully appreciated in Supplemental Video 1.

In VARP, VT was induced after the S2 beat had initially blocked at the isthmus's mouth proximal to the stimulus site (t = 480\,ms after the S2).
As can be seen in Fig.~\ref{fig:Piggy21_VT}B (bottom panel), the wavefront travels around the scar entering the infarct region from a distal site.
The wavefront then propagates back towards the base (t = 750\,ms) exiting through the isthmus from where it reenters the myocardium (t = 1000\,ms).
See Supplemental Video 2 for further details of the block and reentry within Pig 3.

The similarity between the VTs induced by VITA and VARP can be seen at times t = 270\,ms in Fig.~\ref{fig:Piggy21_VT}B and t = 1000\,ms in Fig.~\ref{fig:Piggy21_VT}, respectively.
Both VTs share the same isthmus and have the same chirality, but the VT induced with VARP takes longer to stabilize following the block of the S2.

\subsection{Differences in induced VTs}

While more unique VTs were induced by VITA, inducibility in Pig 4 was only possible with VARP (see Table~\ref{tab:uVTs}).
In contrast to all other induced VTs, the electrical reentry underlying this particular case was anchored around a region of functional block rather than an isthmus.

\begin{figure*}[!t]
    \centering
    \includegraphics[width=\textwidth]{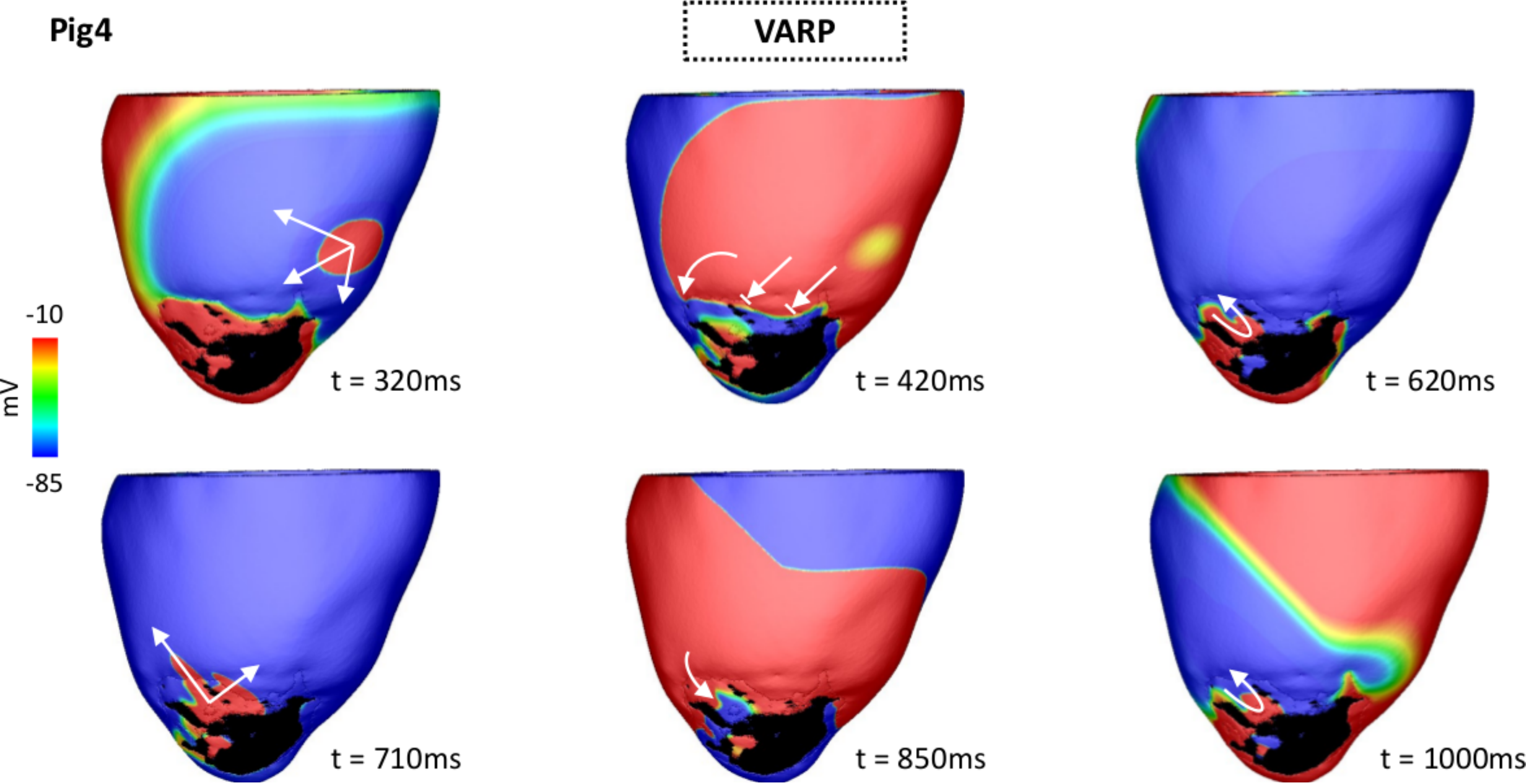}
    \caption{VT initiation due to functional conduction block in Pig 4 heart model using VARP.
             $V_m$ maps at different times show VT induction following the S1-S2 pacing protocol.
             Arrows represent successful propagation.
             Lined arrows represent conduction block.
             The S2 beat blocked at the BZ at t = 420\,ms traveling around the line of conduction block until it reached tissue that had recovered excitability (t = 620\,ms).
             At time t = 710\,ms the wavefront leaves the BZ and enters the LV myocardium.
             It propagates throughout the LV and returns to the BZ forming a reentrant circuit around the region of functional block.
             VT induction in the LV model can be fully appreciated in Supplemental Video 3.
            }
    \label{fig:Piggy23_VT_VARP}
\end{figure*}

Fig.~\ref{fig:Piggy23_VT_VARP} shows $V_\mathrm{m}$ maps following the S2 beat.
The wavefront travels throughout the LV
until encountering a line of block at the infarct BZ (see Fig.~\ref{fig:Piggy21_VT}, panel t = 420\,ms).
As the BZ tissue distal to the S2 stimulus location recovers excitability,
the wavefront can reenter there at t = 620\,ms.
At time t = 710\,ms the wavefront exits the BZ and propagates to the LV.
Note that the reentry sustaining this VT (620\,ms $\leq$ t $\leq$ 1000\,ms) is anchored in the BZ, where CV is slow due to the reduced tissue conductivity in this region.
The conduction block and the evolution of this reentry can be further appreciated in Supplemental Video 3.

Conversely, VT was only inducible in Pig 5 with VITA (see Tab.~\ref{tab:uVTs}).
This VT revolves around a short circuit with a RTT of 78\,ms (data not shown).

\subsection{Ablation targeting}

\begin{figure*}[!t]
    \centering
    \includegraphics[width=\textwidth]{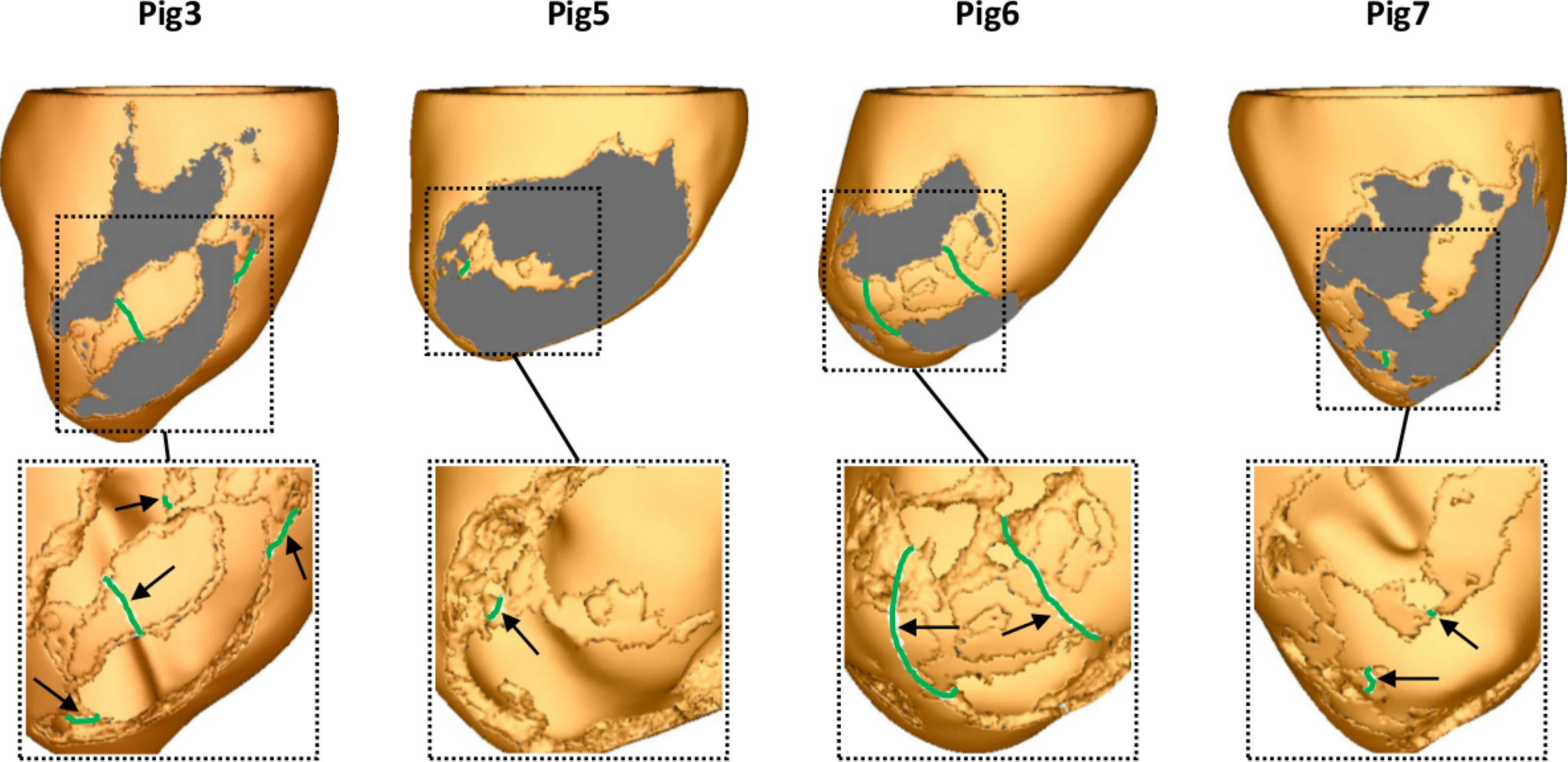}
    \caption{Optimal ablation targets computed by VITA on the ventricular surfaces of Pigs 3, 5, 6 and 7.
             Ablation lesions are shown in green.
             Although ablation targets are displayed on LV epicardial surfaces, each lesion represents a transmural 3D isosurface associated with one exit site selected according to the algorithm in Sect.~\ref{sect:ABL-targeting}.
             Insets show the zoomed-in ablation targets where the scar was removed for visualization purposes.
             Note the presence of ablation lesions under the scar of Pigs 3 and 6.
            }
    \label{fig:Piggies_uVTs_ABL}
\end{figure*}

Once all unique VTs were detected, VITA was employed to detect optimal ablation targets in the virtual hearts.
Computed ablation maps in the LV models in which VT was inducible are shown in Fig.~\ref{fig:Piggies_uVTs_ABL}.
Note that in Pig 3 one region selected for ablation is located inside the isthmus rather than at one of its mouths.
This is a result of a distance field generated by an initiation site
located on the ventricles opposite to the location of the scar.
This led to a collision inside the isthmus which was interpreted, according to the definition given in Sec.\ref{sect:exit_sites},
as an exit site by VITA.
Size and shape of the ablation targets can be better appreciated in the insets (the scars were removed for visualization purposes).
Note the presence of two extra targets located under the scar in Pigs 3.
Although ablation targets are displayed on the epicardial surface,
they are associated with transmural 3D isosurfaces.

Next, \textit{in-silico} ablation was performed to both low and high resolution models of Pig 3 according to the targets in Fig.~\ref{fig:Piggies_uVTs_ABL}.
To assess the success of the virtual ablation, both VITA and VARP inducibility protocols were repeated in low- and high resolution ablated models, respectively.
VTs were no longer inducible in the ablated Pig model.

\subsection{Computational performance}

The computational costs of both approaches applied to Pig 3 are presented in Tables~\ref{tab:VITA_times}-\ref{tab:VARP_times}.
Execution times associated with single as well as multi-site pacing are shown.
Execution times refer to the amount of time required by VITA on a 12-core desktop machine and by VARP on 4 nodes (256 cores) of the HPC used in this study (see Sect.~\ref{sect:comput_environment} for details).
CPU time is given by the execution time divided by the total number of cores used in the simulations.
When pacing from only one location, VITA could induce VTs, detect ablation targets and simulate ECGs in 4\,min on a desktop, while VARP required 4\,h of simulation time on a HPC machine.
Multi-site pacing increased execution times to about 48\,min and 68\,h:34\,min using VITA and VARP, respectively.
Considering CPU time, VITA was, on average, 1\,564 times faster than VARP.

\begin{table}[!t]
    \centering
    \caption{Execution times using VITA on down-sampled models.}
    \begin{tabular}{|c|c|c|c|}
        \hline
        \textbf{Pacing sites}                      & \textbf{Module} & \textbf{Desktop time} & \textbf{CPU time} \\ \hline
        \multicolumn{ 1}{|c|}{\multirow{4}{*}{1}}  & VT inducing     & 0\,h:02\,min:49\,s    & \multicolumn{ 1}{c|}{\multirow{4}{*}{0\,h:47\,min:48\,s}} \\ \cline{ 2- 3}
        \multicolumn{ 1}{|c|}{}                    & Ablation        & 0\,h:00\,min:09\,s    & \multicolumn{ 1}{c|}{} \\ \cline{ 2- 3}
        \multicolumn{ 1}{|c|}{}                    & ECG             & 0\,h:01\,min:01\,s    & \multicolumn{ 1}{c|}{} \\ \cline{ 2- 3}
        \multicolumn{ 1}{|c|}{}                    & Total           & 0\,h:03\,min:59\,s    & \multicolumn{ 1}{c|}{} \\ \hline
        \multicolumn{ 1}{|c|}{\multirow{4}{*}{17}} & VT inducing     & 0\,h:44\,min:34\,s    & \multicolumn{ 1}{c|}{\multirow{4}{*}{9\,h:35\,min:00\,s}} \\ \cline{ 2- 3}
        \multicolumn{ 1}{|c|}{}                    & Ablation        & 0\,h:01\,min:14\,s    & \multicolumn{ 1}{c|}{} \\ \cline{ 2- 3}
        \multicolumn{ 1}{|c|}{}                    & ECG             & 0\,h:02\,min:07\,s    & \multicolumn{ 1}{c|}{} \\ \cline{ 2- 3}
        \multicolumn{ 1}{|c|}{}                    & Total           & 0\,h:47\,min:55\,s    & \multicolumn{ 1}{c|}{} \\ \hline
    \end{tabular}
    \label{tab:VITA_times}
\end{table}

\begin{table}[!t]
    \centering
    \caption{Execution times using VARP on high resolution models.}
    \begin{tabular}{|c|c|c|}
        \hline
        \textbf{Pacing sites} & \textbf{HPC time} & \textbf{CPU time} \\ \hline
        1  & 04\,h:02\,min:02\,s & 1\,032\,h:40\,min:32\,s  \\ \hline
        17 & 68\,h:34\,min:32\,s & 17\,555\,h:20\,min:32\,s \\ \hline
    \end{tabular}
    \label{tab:VARP_times}
\end{table}

\subsection{Sensitivity analysis}

Sensitivity analysis was done to key parameters of VITA: distance between isosurfaces ($D_{\mathrm{iso}}$), RTT and the anisotropic CVs used as parameters for the Eikonal model (Eq.~\ref{eq:Eikonal}).

\subsubsection{Changes in $D_{\mathrm{iso}}$}

Table.~\ref{tab:sensitivity_STEP} shows the number of VTs detected by VITA when $D_{\mathrm{iso}}$ is decreased from 1.0\,mm (default value) to 0.5\,mm and increased to 5.0\,mm.
As in Table~\ref{tab:uVTs}, VTs were induced in Pigs 3, 5, 6 and 7.
The change in number of VTs was inversely related to changes in $D_{\mathrm{iso}}$.
The number of unique VTs did not follow this trend, decreasing in Pig 3 while increasing in Pig 6.

\begin{figure*}[!t]
    \centering
    \includegraphics[width=\textwidth]{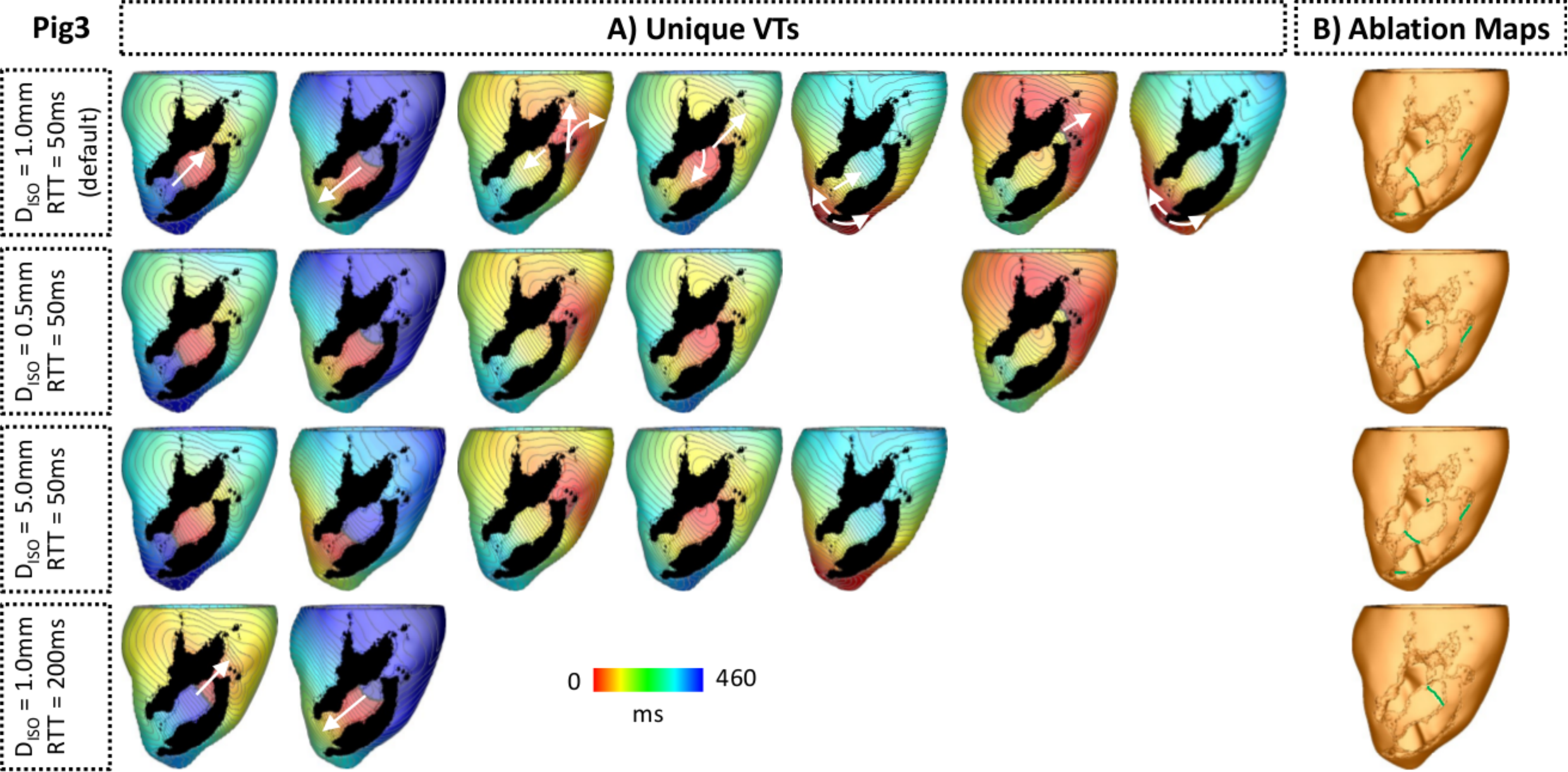}
    \caption{Effects of $D_{\mathrm{iso}}$ and RTT on number of unique VTs and ablation targets in Pig 3.
             A: Unique VTs detected by VITA after applying the multi-site pacing in Pig 3 for each value of $D_{\mathrm{iso}}$ and RTT under investigation.
             B: Optimal ablation maps generated by VITA based on exit sites associated with the unique VTs in A.
             The scar was removed to allow visualization of lesions located in the endocardium.
            }
    \label{fig:Pig21_sensitivity}
\end{figure*}

The effects of changes in $D_{\mathrm{iso}}$ on ablation targets computed within Pig 3 are summarized in Fig.~\ref{fig:Pig21_sensitivity}.
Although changes in $D_{\mathrm{iso}}$ affected the number of unique VTs, the ablation targets remained similar (compare panels Fig.~\ref{fig:Pig21_sensitivity} in B) for constant RTT = 50\,ms).
However, two VTs associated with the presence of a small isthmus near the apex were not detected when $D_{\mathrm{iso}}$ was decreased to 0.5\,mm.
As a result, the more apical ablation target seen on maps computed with $D_{\mathrm{iso}}$ = 1.0\,mm and $D_{\mathrm{iso}}$ = 5.0\,mm is not seen for case of  $D_{\mathrm{iso}}$ = 0.5\,mm.
The major effect of changes in $D_{\mathrm{iso}}$ was observed in execution times.
Decreasing $D_{\mathrm{iso}}$ from 1.0\,mm to 0.5\,mm led to a 2-fold increase in the time to perform the VT inducing module in VITA: from 44\,min:34\,s (Tab.~\ref{tab:VITA_times}) to 01\,h:29\,min:20\,s.
Increasing $D_{\mathrm{iso}}$ to 5.0\,mm shortened the VT detection time to 10\,min:58\,s.

\begin{table}[!t]
    \centering
    \caption{Sensitivity analysis of VITA in respect of changes to distance between isosurfaces ($D_{\mathrm{iso}}$).
             Numbers in parenthesis represent changes in relation to the results obtained with VITA in Tab.~\ref{tab:uVTs}}
    \begin{tabular}{|c|c|c|c|c|}
        \hline
        \multicolumn{ 1}{|c|}{\multirow{2}{*}{\textbf{Pig}}} & \multicolumn{ 2}{c|}{\textbf{Induced VTs}} & \multicolumn{ 2}{c|}{\textbf{Unique VTs}} \\ \cline{ 2- 5}
        \multicolumn{ 1}{|c|}{} & \textbf{0.5\,mm} & \textbf{5.0\,mm} & \textbf{0.5\,mm} & \textbf{5.0\,mm} \\ \hline
        1 & 0   (0)   & 0  (0)   & 0 (0)  & 0 (0)  \\ \hline
        2 & 0   (0)   & 0  (0)   & 0 (0)  & 0 (0)  \\ \hline
        3 & 75  (+29) & 33 (-13) & 5 (-2) & 5 (-2) \\ \hline
        4 & 0   (0)   & 0  (0)   & 0 (0)  & 0 (0)  \\ \hline
        5 & 9   (+4)  & 2  (-3)  & 1 (0)  & 1 (0)  \\ \hline
        6 & 166 (+69) & 59 (-38) & 6 (+1) & 8 (+3) \\ \hline
        7 & 98  (+23) & 48 (-27) & 6 (0)  & 5 (-1) \\ \hline
    \end{tabular}
    \label{tab:sensitivity_STEP}
\end{table}

\subsubsection{Changes in RTT}

Effects of increasing RTT from 50\,ms to 200\,ms are presented in Table.~\ref{tab:sensitivity_RTT}.
Overall, a longer RTT led to fewer detected VTs.
Note that increasing RTT resulted in no VT being detected in Pig 4.
As shown in Fig.~\ref{fig:Pig21_sensitivity}, a longer RTT reduced the number of unique VTs detected by VITA as these are associated with smaller reentrant pathways.
Execution times were not affected by RTT.

\begin{table}[!t]
    \centering
    \caption{Sensitivity analysis of VITA in respect of changes to RTT from 50\,ms to 200\,ms.
             Numbers in parenthesis represent changes in relation to the results obtained with VITA in Tab.~\ref{tab:uVTs}}
    \begin{tabular}{|c|c|c|c|c|}
        \hline
        \textbf{Pig} & \textbf{Induced VTs} & \textbf{Unique VTs} \\ \hline
        1 & 0  (0)   & 0 (0)  \\ \hline
        2 & 0  (0)   & 0 (0)  \\ \hline
        3 & 23 (-23) & 2 (-5) \\ \hline
        4 & 0  (0)   & 0 (0)  \\ \hline
        5 & 0  (-5)  & 0 (-1) \\ \hline
        6 & 25 (-72) & 6 (+1) \\ \hline
        7 & 26 (-49) & 2 (-4) \\ \hline
    \end{tabular}
    \label{tab:sensitivity_RTT}
\end{table}

The effects of a longer RTT on unique VTs and ablation targets computed within Pig 3 can be seen in Fig.~\ref{fig:Pig21_sensitivity}.
Note that the location of the unidirectional block inside the isthmus shifted by changing RTT.

\subsubsection{Changes in CV}

Finally, CVs in the longitudinal and transverse fiber directions were modified to $\pm$25\% of the default value.
The results of the sensitivity analysis are summarized in  Table~\ref{tab:sensitivity_CV}.
Changes in CV had a slight effect on the number of VTs.
The largest change in induced VTs was observed in Pig 7, where the number of VTs increased from 75 (Table~\ref{tab:uVTs}) to 81 for CV$_{-25\%}$, but remained the same for CV$_{+25\%}$.
However, this increase was not reflected in the number of unique VTs which remained the same as in the control case.
Altogether, the number of unique VTs was marginally changed by increasing or decreasing CV in all hearts.

\begin{table}[!t]
    \centering
    \caption{Sensitivity analysis of VITA in respect of changes to CV.
    Numbers in parenthesis represent changes in relation to the results obtained with VITA in Tab.~\ref{tab:uVTs}}
    \begin{tabular}{|c|c|c|c|c|}
        \hline
        \multicolumn{ 1}{|c|}{\multirow{2}{*}{\textbf{Pig}}} & \multicolumn{ 2}{c|}{\textbf{Induced VTs}} & \multicolumn{ 2}{c|}{\textbf{Unique VTs}} \\ \cline{ 2- 5}
        \multicolumn{ 1}{|c|}{} & \textbf{CV$_{-25\%}$} & \textbf{CV$_{+25\%}$} & \textbf{CV$_{-25\%}$} & \textbf{CV$_{+25\%}$} \\ \hline
        1 & 0  (0)  & 0  (0) & 0 (0)  & 0 (0)  \\ \hline
        2 & 0  (0)  & 0  (0) & 0 (0)  & 0 (0)  \\ \hline
        3 & 46 (0)  & 46 (0) & 6 (-1) & 7 (0)  \\ \hline
        4 & 0  (0)  & 0  (0) & 0 (0)  & 0 (0)  \\ \hline
        5 & 5  (0)  & 0  (0) & 1 (0)  & 0 (-1) \\ \hline
        6 & 92 (-5) & 97 (0) & 7 (+2) & 5 (0)  \\ \hline
        7 & 81 (+6) & 75 (0) & 6 (0)  & 6 (0)  \\ \hline
    \end{tabular}
    \label{tab:sensitivity_CV}
\end{table}

\section{Discussion}

We presented a novel, rapid and fully automated computational approach which can be applied to post-infarction heart models for the purpose of:
1) locating reentrant circuits maintaining scar-related VTs;
2) simulating AT maps and ECGs of such VTs; and
3) computing targets to guide ablation therapy in near real-time.
Using only standard desktop computational resources, VITA was shown to be capable of detecting and simulating AT maps of VTs as well as automatically, without operator intervention, generating an optimal ablation map in 48\,min.
Our approach also facilitates the fast simulation of EGMs and ECGs of each uncovered VT circuit, offering the opportunity to compare the simulated ECG or EGM signatures to those recorded clinical, either prior to or during catheter ablation therapy and, thus, confirm the electrophysiological correspondence between simulated and clinical VT.

\subsection{Comparison with R-D approaches}

\textit{In-silico} VT induction protocols have proved to be a highly useful and robust way to reliably induce (multiple) complex arrhythmias in the context of structural heart disease.
Such methods have been successfully applied to demonstrate the potential utility of an \textit{in-silico} approach in guiding VT ablation planning~\cite{Prakosa2018} and also arrhythmic risk stratification~\cite{Arevalo2016}.
The majority of VT inducibility studies, particularly for direct use in a clinical workflow~\cite{Arevalo2016,Prakosa2018}, have been based on sophisticated R-D models.
The advantage of R-D is that these models accurately replicate cardiac electrophysiology, including the mechanisms underlying formation and maintenance of arrhythmias, usually through a monodomain representation of tissue-level electrophysiology, along with biophysically-detailed ionic cell models representing sub-cellular dynamics.
Consequently, in the context of personalized clinical modelling, R-D approaches may be considered to constitute a more physiologically-accurate representation of a patient's heart to probe VT inducibility. 
R-D based VT induction protocols have also been widely used by ourselves~\cite{Campos2019B,Monaci2021} (in more abbreviated forms) and others~\cite{Deng2015,Pashakhanloo2018,Lopez-Perez2019,Deng2019B} in more basic mechanistic studies, investigating the properties of arrhythmias and consequential modelling considerations in this regard.

However, R-D approaches suffer from two main disadvantages.
Firstly, the compute costs required to conduct such a large number of expensive monodomain simulations are punitive, rendering inducibility testing a costly endeavor; the subsequent requirement for off-site HPC facilities hampers the use of such approach within a real clinical workflow.
In comparison, VITA required only a fraction of the computational time of the corresponding R-D  approach (VARP) to achieve the same goal of uncovering scar-related VTs in infarcted ventricular models.
Specifically, more than 68\,h:30\,min of computing time was required to implement the full VARP protocol on an HPC cluster (256 CPUs). Using only a fast desktop machine (with 12 CPUs), VITA brought this time down to approximately 48\,min which is compatible for use within a clinical procedure.

Secondly, R-D models intrinsically use high-dimensional parameter vectors.
Such parameters, principally for the ionic cell models, primarily govern tissue excitability and refractoriness, and thus directly influence arrhythmia induction (through unidirectional block) and maintenance (through electrical wavelength).
The vast majority of these parameters used cannot be constrained with currently available clinical data.
Consequently, VT inducibility using R-D approaches has a strong functional component, which is not patient-specific, but instead relies on literature values for parameterization.
In comparison, VITA only detects vulnerable sites associated with the anatomical presence of an isthmus within a non-conducting scar.
The only functional parameter in VITA relates to the RTT (or, equivalently the assigned CV), which may be more easily tuned to match the known VT cycle length, and even directly compared through EGM/ECG simulation with the clinical VT. 
This important aspect of comparison and validation in clinical \textit{in-silico} VT simulation studies with personalized models has so far been lacking~\cite{Prakosa2018}.

In the context of scar-related VT, anatomical isthmuses (with non-conducting scar on both sides~\cite{Deng2019A}) represent the most common substrate.
This is underscored by the recent suggested success of directly using ``de-channeling'' analysis of CMR data to directly locate isthmuses and guide ablation procedures~\cite{Sanchez-Somonte2021,Berruezo2015}.
Consequently, both approaches, VITA and the R-D based VARP, show very similar performance at uncovering unique VTs in each LV model in the virtual cohort (Table~\ref{tab:uVTs}).

Nonetheless, functional block may form an important substrate to sustain VT which cannot be identified, in its current form, by VITA.
Channels formed within a region of functional conduction block, where one entrance to an isthmus continually blocks the reentrant wave as it is refractory at the time of the wave's passage, with the other side formed by non-conducting scar~\cite{Deng2019A}, such as the one in Fig.~\ref{fig:Piggy23_VT_VARP}, could only be detected in full-blown R-D simulations, as used with VARP.
However, the appearance of such regions of functional block depends sensitively on the specific electrophysiological properties assigned to the healthy tissue and BZ within the model, which thus control the relative degree of refractoriness between these regions.
Currently, in the VARP approach, electrophysiology is not personalized in any way, relying purely on generic literature values of ionic conductances and tissue conductivities.
Whether these regions of functional block are really present in this particular patient/model is unknown, and thus their identification may be of limited value in a clinical context of a personalized therapy.

\subsection{Comparison with R-E approaches}

Alternatively, VT inducibility has also been presented in personalized models using Eikonal approaches~\cite{Chen2016,Cedilnik2018}.
Such studies, like VITA, benefit from a more computationally efficient methodology, requiring less than 1\,m to simulation a full VT activation map~\cite{Cedilnik2018}; this is similar to the time required by the VT inducibility module of VITA, which runs Eikonal simulations for each exit site detected from a given pacing time (Tab.~\ref{tab:VITA_times}).
However, these studies have not explicitly tested for inducibility, as induction was enforced by a local unidirectional block whose location (and direction) required prior interactive manual identification. 
Subsequent simulation of AT maps of the ensuing VT was compared to electroanatomical mapping data, but was not used for identifying the substrate.
Ablation targeting relied entirely on image analysis, and was not guided by the \textit{in-silico} simulation.
In VITA, the detection of vulnerable regions (distal mouths, \textit{i.e.}, exit sites of isthmuses capable of sustaining VTs) and the respective unidirectional block simulation is done in a fully automated way.

The dynamics of VT circuits detected on down-sampled meshes using the R-E model can also be directly investigated in matching high resolution R-D models.
This is achieved by projecting the electrophysiological tissue state during a VT from a lower resolution onto a high resolution grid using Meshtool~\cite{Neic2020},
and by adjusting conductivities for the high resolution grid to match conduction velocities~\cite{MendoncaCosta2013}.
Unlike simulations within the low resolution meshes, tissue conductivities in high resolution models, such as the one in Fig.~\ref{fig:Piggy21_VT}A, may not need to be altered to match prescribed velocities.
However, R-D simulations within high resolution models come with a higher computational cost (see Table~\ref{tab:VARP_times}).

\subsection{Automated ablation target identification}

In this study, VTs were induced using a similar pacing protocol to VARP studies~\cite{Arevalo2016,Prakosa2018}.
However, ablation maps were only automatically computed when using VITA
as the minimum cut algorithm used in VARP studies was not available to us~\cite{Zahid2016,Prakosa2018}.
In principle, the VITA-based ablation target identification could be readily applied
to replace the minimum cut approach in VARP studies
by using activation maps generated by the VARP protocols.
However, to demonstrate the ability of the set of VITA ablation lesions
to render virtual hearts VT-free the VARP protocol was applied to Pig 3 model.
VITA was also repeated on the ablated model to investigate whether new channels were formed as a result of the ablation.
The LV model was ablated according to the targets identified by VITA (maps in Fig.~\ref{fig:Piggies_uVTs_ABL}).
VT was no longer inducible in the newly ablated model, demonstrating the ability of our approach to eliminate structural substrates sustaining VT.

\subsection{Parameter sensitivity}

Owing to its computational efficiency, VITA is ideally suitable
for exploring the sensitivity of its output (unique VTs, corresponding circuits and ablation targets)
to key input parameters (both of the VITA algorithm itself and related the electrophysiological properties).

\subsubsection{$D_{\mathrm{iso}}$}

While the number of VTs changed with $D_{\mathrm{iso}}$ (Table.~\ref{tab:sensitivity_STEP}),
the main impact of increasing $D_{\mathrm{iso}}$ was the reduction in computational time.
A larger $D_{\mathrm{iso}}$ results in less split isosurfaces inside the isthmus speeding up the algorithm.
The computational cost associated with the VT induction step of VITA in Pig 3
decreased by about a factor of 4 (from 44\,min:34\,s to 10\,min:58\,s),
as $D_{\mathrm{iso}}$ increased from 0.5\,mm to 5\,mm.

\subsubsection{RTT}

The VT cycle length is an estimation of the time needed by the wavefront
to complete one full cycle around a reentrant pathway.
VTs with cycle lengths ranging from 127 $\pm$ 72\,ms to 365\,ms have been reported in the literature~\cite{Volosin1991,Nishimura2021}.
This time is taken into account by adjusting RTT in VITA (see Fig.~\ref{fig:H2C_uVTs_STIM}).
This can be done also in a patient-specific manner,
guided by pre-operative recordings of the clinical VT, where available.
A longer RTT of 200\,ms resulted in less VTs being detected
as reentrant circuits associated with shorter path lengths were disregarded by the algorithm
(see Table.~\ref{tab:sensitivity_RTT}).
Indeed, the shorter circuit associated with the only unique VT detected in Pig 5 was ignored
when RTT was prolonged to 200\,ms.
Interestingly, the use of a longer RTT resulted in a similar number of unique VTs by VITA (Table.~\ref{tab:sensitivity_RTT}) and VARP (Table~\ref{tab:uVTs}).
This is likely due to the longer APD in the BZ that prevented reentry of wavefronts at isthmuses
associated with shorter circuits~\cite{Campos2019B}.

The change in the number of unique VTs when altering RTT also affected the predicted ablation map
as shown in Fig.~\ref{fig:Pig21_sensitivity}.
RTT may vary from patient to patient as well as during the ablation procedure.
By considering a very short RTTs (50\,ms), VITA was capable of unmasking all possible circuits,
even those with RTT smaller than typical VT cycle lengths.
Such small circuits might not be able to sustain clinically relevant VTs at a specific stage of the disease.
However, remodelling process that progressively slow CV or shorten APD within the isthmus
could reduce the wavelength, allowing these channels, potentially, to sustain VTs~\cite{MendoncaCosta2018}.
Thus, by detecting isthmuses associated with a short RTT,
VITA may potentially eliminate the need for repeated ablations due to VT recurrence.

\subsubsection{EP properties}

The computational efficiency of VITA allows for the assessment of different scenarios
such as the slowing of CV or the shortening of APD, that could uncover VT circuits
that may form only for short wavelengths.
We found that inducibility in all models was similar
despite the relatively large changes in CV applied (+/-25\%).
Overall, changes to CV had the expected effect on VT inducibility:
decreasing CV shortened the wavelength of the AP, allowing to fit an entire AP
inside smaller reentrant circuit (more unique VTs).
The increase in CV had the opposite effect, leading to fewer VTs induced
as shorter circuits were not able any more to fit an entire AP wave with a minimum excitable gap.
These findings are in accordance with a previous study using full-blown R-D simulations,
which demonstrated that VT localization is fairly insensitive to EP variability,
suggesting that infarct anatomy plays the major role
in formation and maintenance of scar-related VTs~\cite{Deng2019A}.

\subsection{Utility in clinical applications }

VITA automatically identified exit sites defined here as distal mouths of channel isthmuses within the scar.
Exits rather than entries were chosen here because they identify the point of activation of the myocardium and therefore serve as an important guide for catheter ablation therapy.
We have shown that our novel approach can faithfully reproduce VT circuits within infarcted models, similar to previously used approaches relying on R-D models (VARP).
For these their clinical utility in stratifying arrhythmic risk or in guiding catheter ablation targets has been demonstrated already with real patient models~\cite{Arevalo2016,Prakosa2018}.
However, the computational costs of VARP-like approaches relying on the availability of HPC resources
is almost prohibitively expensive for clinical applications.
An important advance of our method in this regard is the ability to yield comparable results
at a fraction of the costs.
VITA is orders of magnitudes more efficient and can be executed
with standard desktop computing power.
The efficiency of VITA opens up perspectives of optimizing VT ablation,
either off line pre-operatively or, potentially, online during the ablation procedure itself.

A major issue in VT ablation is identifying the clinical VT responsible for the patient's presentation.
To trust computational models to be used in therapy optimization evidence is needed
to demonstrate that modelled VT circuits correspond closely to the clinical VT observed in a given patient.
VITA as implemented in here based on a R-E model offers detailed electrophysiology simulations
with close to real-time performance~\cite{Gillette2021},
allowing to produce all data relevant for comparing with EAM systems,
including device and catheter EGMs as well as ECGs or body surface potential maps.
ECGs and EGMs in our study were computed based on the $\phi_e$-recovery method
that relies on simplifying assumptions~\cite{Plonsey1998}.
This could be replaced with the computationally more efficient lead field method,
that yields EGMs of maximum biophysical fidelity~\cite{Potse2018,Gillette2021}.
This combined approach facilitates the rapid and accurate simulation of EGM recordings
associated to VTs as they are measured \textit{e.g.}\ by implantable cardioverter defibrillators (ICDs)
which the vast majority of ablation patients have \textit{in-situ},
as performed in our own recent study~\cite{Monaci2021}.
Such an ability provides the potential for direct comparison with pre-operative recordings
of the clinical VT from patient ICDs (or, alternatively, ECG measurements, if available), allowing a direct matching of simulated VT circuit (and ablation targets) with the clinical VT of a given patient.
Close matching of EGM and ECG wave forms can be considered an important pathways
towards enhancing credibility and clinical confidence in simulated predictions.

Due to its speed, VITA can not only be used to directly match the clinical VT cycle length
(if known, for example, from ICD recordings) through the RTT parameter,
but also used to uncover any potential VT circuit and ablation target
by varying this parameter and repeating the analysis.
This may not only facilitate the targeting of the clinical VT,
but also, potentially, reducing the need for re-do procedures
by ensuring that other possible substrates are eliminated at the time of initial ablation.

\subsection{Clinical imaging and model construction}

It is important to note that the accuracy and overall utility of our method, in a similar manner to all previous works in this field, depends upon the fidelity of the clinical imaging and subsequent segmentation performed to construct the computational model.
Here, we utilized 1\,mm isotropic CMR, which is higher resolution than most current clinical acquisitions.
Such a resolution is important to robustly identify the structural anatomy of critical isthmuses through the scar which represented the substrate for sustaining the VT.
However, previous simulation studies have used clinical CMR acquired at similar in-plane resolution (1-2\,mm), but much lower slice spacing of 8-12\,mm~\cite{Deng2015,Deng2019B,Prakosa2018}.
These studies suggested that the VTs uncovered and their ensuing dynamics are not significantly affected by this.
This is in stark contrast to our own recent analysis which indicated
that a lower out of plane resolution significantly alters VT dynamics (unpublished data).
Nonetheless, imaging capabilities and sequences are continually improving.
With a recent drive towards fully 3D CMR nearly it is anticipated
that higher resolution 3D MRI will become standard clinical practice~\cite{Milotta2021},
or, alternatively, higher resolution CT yielding image stacks of sub-millimeter resolution
could be used for anatomical reconstruction.
Thus, issues surrounding adequate image resolutions for model construction
may well be obsolete in the foreseeable future.
Beyond anatomy, image analysis to delineate scar and BZ from viable healthy myocardium
is even more important.
Validation and optimization of segmentation approaches are also rapidly evolving for both CMR~\cite{Karim2013,Razeghi2020} and CT~\cite{Yamashita2016,Cedilnik2018},
which will facilitate, the robust and accurate model generation
from higher resolution clinical images.

In addition, the majority of patients undergoing VT ablation therapy have ICDs \textit{in-situ}
prior to ablation procedures, which can cause artifacts in CMR images.
Such artifacts can be variable, rendering segmentation for an accurate model construction
more challenging but still feasible~\cite{Prakosa2018}.
The evolution of novel wideband sequences is addressing the need for suppressing artifacts in patients with cardiac implanted electronic devices~\cite{Rashid2014,Ranjan2015,Rashid2016}.
Finally, refinements in contrast and dual-energy CT imaging,
along with extra-cellular volume reconstructions, also promises potential utility
for deriving robust information about scar anatomy for computational infarct model constructions.
With CT imaging device artifacts are less prominent
and overall image resolution is intrinsically higher relative to CMR.

\subsection{Limitations of VITA}

As discussed previously, one of the main limitations of VITA is that its outcome relies on the quality of the segmented images.
However, this shortcoming will be lessened by the continuous advance in both data acquisition and image analysis to delineate scar and BZ from healthy tissue.
Also, in its current form VITA is not suitable for studying functional VTs.
Functional conduction block results from dispersion of repolarization.
Such detail cannot yet be robustly acquired by current EAM systems, which renders representation of such electrophysiological detail in a patient-specific manner not possible in this approach or any other.
Furthermore, while VITA allows for fast simulation of ECGs and EGMs, a comparison between these and the dynamics of the clinical VT in a patient-specific manner has not been performed yet.
Such direct comparison may also require patient-specific tunning of CV and RTT in VITA to accurately match the presenting VT.
Finally, anatomical structure may develop and change during the course of disease progression (structural remodelling - from days to months), emphasizing the requirement of imaging used for model construction just prior to ablation therapy planning for optimal model-based predictions.

\section{Conclusion}

This study presented a novel automated computational tool for induction and treatment of scar-related VTs in near-real time.
VITA was shown to be capable of rapidly detecting VT exit sites, simulating AT maps and ECGs as well as to output an optimal set of targets for catheter ablation therapy.
The rapid algorithmic approach proposed here, combined with personalization techniques for building digital twin models that facilitate the detailed replication of electrophysiological recordings, may be used as a complementary modality aiding in clinical decision making in the treatment of post-infarction VTs.

\section*{Acknowledgments}

This research was supported by the National Institute for Health Research (NIHR) Clinical Research Facility at Guy's and St. Thomas' National Health Service (NHS) Foundation Trust and NIHR Biomedical Research Centre based at Guy's and St. Thomas' NHS Foundation Trust and King's College London.
The views expressed are those of the authors and not necessarily those of the NHS, the NIHR or the Department of Health.
Open access for this article was funded by King's College London.
This research was funded in whole, or in part, by the Wellcome Trust [213342/Z/18/Z]. For the purpose of open access, the author has applied a CC BY public copyright license to any Author Accepted Manuscript version arising from this submission.
FOC and MJB acknowledge the support of the British Heart Foundation through Project Grant PG/16/81/32441.
JW acknowledges the support of the the Welcome Engineering and Physical Sciences Research Council (EPSRC) Centre for Medical Engineering at King's College London (WT 203148/Z/16/Z), the Health Innovation Challenge Fund (HICF-R10-698), a parallel funding partnership between the Department of Health and the Welcome Trust, and the EPSRC grant (EP/R010935/1).
MJB acknowledges the support of the UK Medical Research Council through a New Investigator Research Grant number MR/N011007/1.
GP acknowledges the support of the grants I2760-B30 from the Austrian Science Fund (FWF),
BioTechMed-Graz Flagship award ILearnHeart, and MedalCare 18HLT07 from the EU.

\bibliographystyle{rusnat}
\bibliography{refs}

\section*{Supplementary material}

\textbf{Supplemental video 1:}
Induced VT within Pig 3 model using VITA.
Transmembrane voltage ($V_m$) of the reentrant wave in the high resolution (same scale as Fig.~\ref{fig:Piggy21_VT}).
$V_m$ and other state variables were mapped from the down-sampled (R-E) onto the high resolution (R-D) model.

\textbf{Supplemental video 2:}
Induced VT within Pig 3 model using VARP.
Conduction block and reentry following a premature S2 beat delivered near the base of the LV model.
Transmembrane voltage ($V_m$) of the premature S2 (coupling interval of 250\,ms) that blocked at the isthmus and degenerated into reentry (same scale as Fig.~\ref{fig:Piggy21_VT}).

\textbf{Supplemental video 3:}
Induced VT within Pig 4 model using VARP.
Functional block and reentry following a premature S2 beat delivered at a mid point between the apex and base of the LV model.
Transmembrane voltage ($V_m$) of the premature S2 (coupling interval of 250\,ms) that blocked at infarct BZ and degenerated into reentry (same scale as Fig.~\ref{fig:Piggy23_VT_VARP}).

\end{document}